\documentclass[twocolumn]{aastex631}

\usepackage[T1]{fontenc}
\usepackage{graphicx}
\usepackage{amsmath}
\usepackage{natbib}
\usepackage{longtable}
\usepackage{xcolor}

\shorttitle{Accretion Rate of Young Intermediate Mass Stars}
\shortauthors{Brittain et al.}

\begin{document}

\title{Evolution of the Accretion Rate of Young Intermediate Mass Stars: Implications for Disk Evolution and Planet Formation}

\author[0000-0001-5638-1330]{Sean D. Brittain}
\affil{Department of Physics and Astronomy \\ 
Clemson University \\
Clemson, SC 29634-0978, USA}

\author[0000-0002-5860-6043]{Joshua W. Kern}
\affil{Department of Physics and Astronomy \\ 
Clemson University \\
Clemson, SC 29634-0978, USA}

\author[0000-0002-6251-0108]{Gwendolyn Meeus}
\affil{Universidad Autonoma de Madrid  \\
Madrid, Spain}

\author[0000-0001-7703-3992]{Ren\'e D. Oudmaijer}
\affil{Royal Observatory of Belgium \\ Ringlaan 3 \\ 1180 Brussels,  Belgium }
\affil{School of Physics and Astronomy \\
University of Leeds \\
Leeds LS2 9JT, UK}


\begin{abstract}
This work presents a study of the evolution of the stellar accretion rates of pre-main-sequence intermediate-mass stars. We compare the accretion rate of the younger intermediate-mass T Tauri stars (IMTTSs) with the older Herbig stars into which they evolve. We find that the median accretion rate of IMTTSs (1.2$\times$10$^{-8}$~M$_{\odot}$~yr$^{-1}$) is significantly lower than that of Herbig stars (1.9$\times$10$^{-7}$~M$_{\odot}$~yr$^{-1}$). This increase stands in stark contrast with canonical models of disk evolution that predict that the stellar accretion rate declines with age.  We put forward a physically plausible scenario that accounts for the systematic increase of stellar accretion based on the increase of the effective temperature of the stars as they evolve towards the zero-age main sequence. For example, the temperature of a 2M$_{\odot}$ star will increase from 4900~K in the IMTTS phase to 9100~K during the Herbig phase. Thus, the luminosity of the far ultraviolet (FUV) radiation will increase by orders of magnitude. We propose that this increase drives a higher stellar accretion rate.  The scenario we propose to account for the increase in the stellar accretion rate solves the lifetime problem for Herbig disks because the increasing stellar accretion rates require lower initial disk masses to account for present-day disk masses. This work highlights the importance of the role FUV radiation has in driving the accretion rate, predicts a large population of pre-main-sequence non-accreting A stars, and has implications for interpreting disk morphologies that may serve as signposts of embedded gas giant planets in Herbig disks.
\end{abstract}

\section{Introduction}

Herbig stars are pre-main-sequence stars that represent an interesting transition between low-mass stars and high-mass stars (see \citealt{mendigutia2020, Brittain_feb23} for recent reviews). Like their lower mass counterparts, the classical T Tauri stars (CTTSs), they reveal extended disks with structures indicative of ongoing gas giant planet formation \citep{dong_aug18, Janson_feb21,Stapper_feb22}. Magnetohydrodynamic (MHD) simulations of ionized disks have shown that the dynamics of the in situ disk gas are strongly affected by magnetic fields that thread the disk \citep{Bai_may13,Bai_sep11}. In the inner disk, these magnetic fields largely arise from the star at the center of the disk. The energy transport mechanisms of T Tauri stars facilitate convection currents that drive strong, well-ordered magnetic fields on the order of a kilogauss \citep[e.g.,][]{JohnsKrull2007}. In contrast, the higher temperatures of Herbig stars sufficiently ionize the outer layers, allowing for a purely radiative transfer of energy \citep{Villebrun_19}. 

Although Herbig stars lack convection currents, magnetic fields have been measured in roughly 10\% of Herbig stars in the range of $\sim$100-1000~G \citep{Alecian_2013}. These fields are likely fossil fields left over from a previous evolutionary phase as an intermediate-mass T Tauri star (IMTTS) or the collapsing parent molecular cloud  \citep{Vioque_dec18}.

Due to the dearth of strong, well-ordered stellar magnetic fields among Herbig stars, it was not expected that the magnetospheric accretion paradigm, so successful at characterizing accretion onto CTTSs, would apply to these earlier-type stars. However, several lines of evidence have emerged to the contrary. For example, H$\alpha$ spectropolarimetric observations indicate that the star/disk interface of CTTSs and Herbig Ae stars is similar and consistent with the presence of an inner hole as expected from magnetically mediated accretion \citep{Vink2002, Vink2005, Vinkmodel2005}. An additional line of evidence comes from successfully modeling the hydrogen emission line profiles and the Balmer discontinuity of UX Ori using a magnetospheric model \citep{Muzerolle_dec04}. A third line of evidence comes from the observation of high-velocity red-shifted absorption features superimposed on hydrogen emission lines that are indicative of gas in freefall onto these stars \citep{Guimares_oct06}. An early effort to calibrate the line luminosity of Br$\gamma$ to the accretion luminosity of Herbig stars as measured from the Balmer discontinuity found that the calibration was consistent with the relationship found for CTTSs \citep{Donehew_feb11}. Subsequent work applied to a larger sample and more lines found that these relationships extended to Herbig stars as early as B7 \citep{mendigutia2011, Fairlamb_feb17, Wichittanakom_mar20}. 

The relationship between the stellar mass and stellar accretion rate among CTTSs is also found for Herbig stars. \citet{Wichittanakom_mar20} found that this relationship extends to $\sim$4M$_{\odot}$ while the relationship is not as steep for the higher mass stars. Expanding on this, \cite{grant_feb22} performed the largest high-resolution spectroscopic survey of Br$\gamma$ emission from Herbig stars, and confirmed these trends with the same break in the dependence of accretion rate on stellar mass occurring near M$_{\star}$ $\sim$ 4M$_{\odot}$. These studies largely confirm evidence from H$\alpha$ spectropolarimetry indicating that the star/disk interface changes for stars with masses $\gtrsim$4~M$_{\odot}$ \citep{Vink2002, Vink2005, Vinkmodel2005, Vink2015, Ababakr2016, Ababakr2017}.

Large surveys of the accretion rates of Herbig stars indicate that the median accretion rate of sources with $\rm M_{\star}\leq4M_{\sun}$ is $\sim$1.9 $\times$ 10$^{-7}$ M$_{\odot}$ yr$^{-1}$ \citep{Fairlamb_feb17, Fairlamb_oct15, Wichittanakom_mar20}. This is markedly higher than the median accretion rate of CTTSs in Taurus which is 1.3 $\times$ 10$^{-8}$ M$_{\odot}$ yr$^{-1}$ \citep{Najita2015}. 

\cite{Wichittanakom_mar20} noted that the relationship of stellar accretion rates with mass is also confounded with the age of the star. To explore the dependency of the stellar accretion rate on the age of the star, these authors selected a narrow range of stellar masses (2.0-2.5 M$_{\odot}$) to minimize the mass dependency of the stellar accretion rate and determined that the accretion rate declines as Age$^{-1.95\pm0.49}$. If stellar accretion rates decline over time, which is expected for viscous accretion models with a constant $\alpha$ \citep{Hartmann_98}, then we should expect the evolutionary precursors of Herbig stars (i.e., the IMTTSs) to have accretion rates that are systematically higher than those of Herbig stars. 

There are few studies dedicated to the study of IMTTSs. These can be defined as stars with $\rm 1.5~M_{\sun} \leq M_{\star} \leq 4.0~M_{\sun}$ and $\rm T_{eff}<7,200~K$ (i.e., stars with a spectral type of F0 or later). One important exception is a study of the accretion rate of nine IMTTSs \citep{Calvet_sep04}. The average stellar accretion rate of these sources was found to be comparable to their lower-mass counterparts, the T Tauri stars with $\rm M_{\star} \leq 1.5~M_{\sun}$ (\.{M}=2.8 $\times$ 10$^{-8}$ M$_{\odot}$ yr$^{-1}$). More recently, \citet{Valegard_apr21} determined the stellar parameters of a large sample of IMTTSs (with mass limits 1.5M$_{\sun}$\textless M$_{\star}$\textless 3.5M$_{\sun}$) within 500~pc, which presents a new opportunity to expand the accretion statistics of this subclass of stars.

Here we extend the work of \citet{Calvet_sep04} by determining the accretion rate of an additional 33 IMTTSs. We show that the accretion rates of IMTTSs from the earlier study are representative of a much larger sample and about an order of magnitude less than the accretion rates of Herbig stars. We propose a solution for this counterintuitive result and discuss how it may account for the apparent discrepancy between the high accretion rate and relatively low disk mass of Herbig stars.

\section{Sample} \label{sec:samp}
The sample of IMTTSs included in this study was drawn from \citet{Valegard_apr21} for which a uniform set of stellar parameters had been determined. To restrict our sample to stars with spectral type F0 and later, we selected the stars with $\rm T_{eff}<7,200~K$ for a total of 47 stars. We found the requisite data in the literature to calculate the stellar accretion rate for 42 of the stars in this sample (Table \ref{t:sample}). To keep the calculation as uniform as possible, the accretion rate was determined from the luminosity of H$\alpha$ for 38 sources. There are four sources for which we did not find a measurement of H$\alpha$ in the literature, and we used alternative methods to determine the stellar accretion rate. The accretion luminosities of two of these were adopted from values measured from the UV excess. One source was inferred from the luminosity of Pa$\beta$, and the other source was inferred from the Br$\gamma$ luminosity. These are noted in Table \ref{t:sample}.

To convert the equivalent width H$\alpha$ to an accretion rate, we first inferred the intrinsic equivalent width of the line by correcting for photospheric absorption using model stellar atmospheres from \citet{Coelho2014}. We do not attempt to correct for the veiling of the H$\alpha$ photospheric line. Doing so would lower the accretion rates we infer and increase the disparity between the accretion rate of IMTTSs and Herbig stars. The line flux was scaled using the Gaia DR2 photometry to remain consistent with the \citet{Valegard_apr21}. We corrected for reddening based on the color excess reported by \citet{Valegard_apr21}, $\rm R_V = 3.1$, and the reddening law from \citet{CCM1989}. The accretion luminosity was inferred from the luminosity of the H$\alpha$ line using the relationship determined by \citet{Alcala_apr17}. We adopted this relationship, calibrated against T Tauri stars rather than the relationship calibrated against Herbig stars by \citet{Fairlamb_feb17} because we expect the accretion geometry and magnetic field strengths of IMTTSs to be more similar to CTTSs than to the radiative Herbig stars \citep{Villebrun_19}. The stellar accretion rate is proportional to the accretion luminosity so that

\begin{equation}
\dot{M} = \frac{L_{acc} R_\star} {GM_\star}, 
\label{eq:mdot}
\end{equation}

\noindent where stellar parameters were taken from \citet{Valegard_apr21}. The median of the stellar accretion rate of our IMTTS sample is $1.2\times10^{-8}~M_{\sun}~yr^{-1}$.

\begin{deluxetable*}{lccclccclc}
\tabletypesize{\scriptsize}
\caption{IMTTS Sample}
\tablehead{\colhead{Name}	&	\colhead{$\rm T_{eff}$}	&	\colhead{L$_{\star}$/L$_{\sun}$} &	\colhead{R}	            &	\colhead{$\rm W_{H\alpha}$}	 &	\colhead{$\rm log(F_{H\alpha})$}	&	\colhead{log(L$\rm _{Acc})$} &	\colhead{log(\.{M})}	        &	\colhead{References}		\\
	                       &	\colhead{K}	    &                                &	\colhead{dereddened}	&	\colhead{\AA}	             &	\colhead{$\rm erg~s^{-1}~cm^{-2}$}	&	\colhead{$\rm L_{\sun}$}	 &	\colhead{M$_{\sun}$ yr$^{-1}$}	&		}
\startdata		
AK Sco       	    &	6250	&	6.94	&	8.6	    &	-4.7	&	-11.1	&	-0.92	&	-8.2	&	1	\\
Ass ChaT2-21 [a]	&	5660	&	10.37	&	8.3	    &	-0.1	&	-13.1	&	-2.92	&	-10.2	&	2	\\
Ass ChaT2-54 [b]	&	5260	&	5.03	&	9.4	    &	\nodata	&	\nodata	&	-2.08	&	-9.4	&	3	\\
BE Ori [a]	        &	5720	&	7.95	&	10.2	&	-33.7	&	-11.3	&	-0.13	&	-7.4	&	4	\\
Brun 252 [c]     	&	5890	&	7.16	&	10.4	&	-4.0	&	-12.3	&	-1.22	&	-8.5	&	5	\\
Brun 555     	    &	5040	&	11.37	&	10.3	&	-3.5	&	-12.3	&	-1.09	&	-8.3	&	6	\\
Brun 656 [d]	    &	5770	&	27.4	&	9.3	    &	\nodata	&	\nodata	&	\nodata	&	\nodata	&		\\
BX Ari [d]	        &	5040	&	48.81	&	2.4	    &	\nodata	&	\nodata	&	\nodata	&	\nodata	&		\\
CO Ori       	    &	6030	&	43.82	&	10.1	&	-7.6	&	-11.9	&	-0.74	&	-7.9	&	7	\\
CR Cha       	    &	4800	&	3.72	&	9.6	    &	-30.8	&	-11.1	&	-0.58	&	-7.8	&	7	\\
CV Cha       	    &	5280	&	4.63	&	9.4	    &	-72.6	&	-10.6	&	-0.05	&	-7.4	&	7	\\
DI Cep [e]       	&	5490	&	9.82	&	9.4	    &	\nodata	&	\nodata	&	0.18	&	-7.1	&	8	\\
DI Cha       	    &	5770	&	9.82	&	8.5	    &	-19.4	&	-10.9	&	-0.32	&	-7.6	&	7	\\
EM*SR 21 [f]     	&	5950	&	7.03	&	10.7	&	\nodata	&	\nodata	&	-1.92	&	<-9.2	&	9	\\
EZ Ori       	    &	5830	&	7.54	&	10.5	&	-14.6	&	-11.8	&	-0.62	&	-7.9	&	7	\\
GW Ori       	    &	5700	&	35.47	&	8.6	    &	-31.8	&	-10.7	&	0.60	&	-6.6	&	7	\\
GX Ori [b]       	&	5410	&	3.1	    &	11.8	&	\nodata	&	\nodata	&	-0.08	&	-7.5	&	10	\\
Haro 1-6     	    &	5880	&	12.78	&	10.3	&	-4.5	&	-12.2	&	-2.17	&	-9.4	&	11	\\
HBC 338      	    &	5490	&	5.67	&	9.9	    &	-28.1	&	-11.3	&	-0.33	&	-7.7	&	12	\\
HBC 415      	    &	5770	&	7.48	&	8.5	    &	-7.3	&	-11.3	&	-0.93	&	-8.2	&	6	\\
HBC 442      	    &	6170	&	13.2	&	9.7	    &	-7.3	&	-11.8	&	-0.63	&	-7.9	&	1	\\
HBC 502      	    &	4830	&	10.71	&	10.3	&	-4.5	&	-12.2	&	-1.08	&	-8.2	&	13	\\
HD 135344B   	    &	6640	&	8.09	&	8.0	    &	-9.7	&	-10.8	&	-0.52	&	-7.3	&	1	\\
HD 142527    	    &	6500	&	19.31	&	7.4	    &	-13.1	&	-10.3	&	0.65	&	-6.6	&	1	\\
HD 288313A [d]  	&	5040	&	38.47	&	9.0	    &	\nodata	&	\nodata	&	\nodata	&	\nodata	&		\\
HD 294260    	    &	6115	&	8.74	&	10.2	&	-16.5	&	-11.6	&	-0.40	&	-7.7	&	7	\\
HD 34700     	    &	6060	&	24.34	&	8.8	    &	\nodata	&	-11.3	&	0.40	&	-6.8	&	1	\\
HD 35929        	&	7000	&	80.78	&	8.0	    &	\nodata	&	-10.9	&	0.91	&	-7.3	&	1	\\
HQ Tau       	    &	5280	&	4.63	&	10.1	&	-4.1	&	-12.2	&	-1.98	&	-9.3	&	14	\\
HT Lup  	        &	4830	&	5.95	&	8.7	    &	-6.3	&	-11.4	&	-1.13	&	-8.3	&	15	\\
LkH$\alpha$ 310    	&	5590	&	6.8	    &	12.8	&	-19.4	&	-12.6	&	-1.47	&	-8.8	&	13	\\
LkH$\alpha$ 330    	&	6240	&	14.39	&	9.2	    &	\nodata	&	-11.0	&	0.07	&	-7.2	&	16	\\
PDS 115      	    &	5770	&	4.27	&	10.2	&	-11.8	&	-11.8	&	-0.93	&	-8.3	&	15	\\
PDS 156         	&	5660	&	21.12	&	8.2   	&	-9.6	&	-11.0	&	0.22	&	-7.0	&	15	\\
PDS 277         	&	6720	&	9.68	&	9.8    	&	\nodata	&	-11.5	&	-0.47	&	-7.2	&	1	\\
PR Ori [d]      	&	5170	&	10.83	&	10.1	&	\nodata	&	\nodata	&	\nodata	&	\nodata	&		\\
RY Ori          	&	6120	&	9.01	&	10.0	&	\nodata	&	-11.4	&	-0.24	&	-7.5	&	1	\\
RY Tau            	&	5945	&	11.97	&	8.8 	&	-23.2	&	-10.9	&	-0.68	&	-7.1	&	6	\\
SU Aur          	&	5680	&	12.75	&	8.4   	&	-6.6	&	-11.3	&	-0.95	&	-8.2	&	6	\\
SW Ori          	&	5490	&	3.28	&	11.2	&	-6.7	&	-12.4	&	-1.36	&	-9.0	&	6	\\
T Tau           	&	5700	&	8.88	&	8.0	    &	-74.4	&	-10.1	&	0.30	&	-7.3	&	7	\\
UX Tau A        	&	5490	&	8.91	&	11.1	&	-8.5	&	-12.3	&	-2.19	&	-8.9	&	7	\\
V1044 Ori       	&	5500	&	6.1	    &	10.6	&	-6.3	&	-12.2	&	-1.08	&	-7.7	&	7	\\
V1650 Ori         	&	6160	&	9.53	&	9.9  	&	-14.2	&	-11.5	&	-0.50	&	-8.3	&	17	\\
V2149 Ori [d]   	&	6180	&	35.8	&	8.8    	&	\nodata	&	\nodata	&	\nodata	&	\nodata	&		\\
V395 Cep         	&	5470	&	4.71	&	9.2   	&	-10.2	&	-11.4	&	-0.95	&	-8.8	&	6	\\
V815 Ori        	&	5530	&	5.46	&	10.8	&	-5.2	&	-12.3	&	-1.24	&	-8.7	&	6	\\
\enddata
   \label{t:sample}                   
\tablecomments{
The stellar temperature and luminosity are taken from \citet{Valegard_apr21}. The accretion luminosity is inferred from published values of the H$\alpha$ unless otherwise noted. Notes: [a] The reported equivalent width was corrected for photospheric absorption. [b] The flux of H$\alpha$ was not reported in literature, so the $L_{acc}$ inferred from the UV was used to calculate the accretion rate. 
[c]	The equivalent width of H$\alpha$ was extracted from the spectrum available in Vizier.
[d]	No data was found in the literature. 
[e]	The accretion luminosity was inferred from a published value for Br$\gamma$. 
[f]	The accretion luminosity was inferred from a published value fo Pa$\beta$. 
References: 
[1]	\citealt{Wichittanakom_mar20}
[2]	\citealt{Walter_aug92}
[3]	\citealt{Manara_aug17}
[4]	\citealt{Fang_jul13}
[5]	\citealt{Villebrun_19}
[6]	\citealt{Herbig_88}
[7]	\citealt{reipurth1996}
[8]	\citealt{Eisner2007}
[9]	\citealt{Natta_jun06}
[10]	\citealt{Calvet_sep04}
[11]	\citealt{Jensen_sep09}
[12]	\citealt{Maheswar_apr03}
[13]	\citealt{Flaherty2008}
[14]	\citealt{Simon_nov16}
[15]	\citealt{Gregorio-Hetem_02}
[16]	\citealt{Manara_14}
[17]	\citealt{Rojas_jul08}
}																		
\end{deluxetable*}

\section{Accretion Rates} 
\begin{figure}[ht]
\begin{center}
\includegraphics[trim = 10mm 5mm 10mm 23mm, clip, width=0.75\linewidth, angle=90]{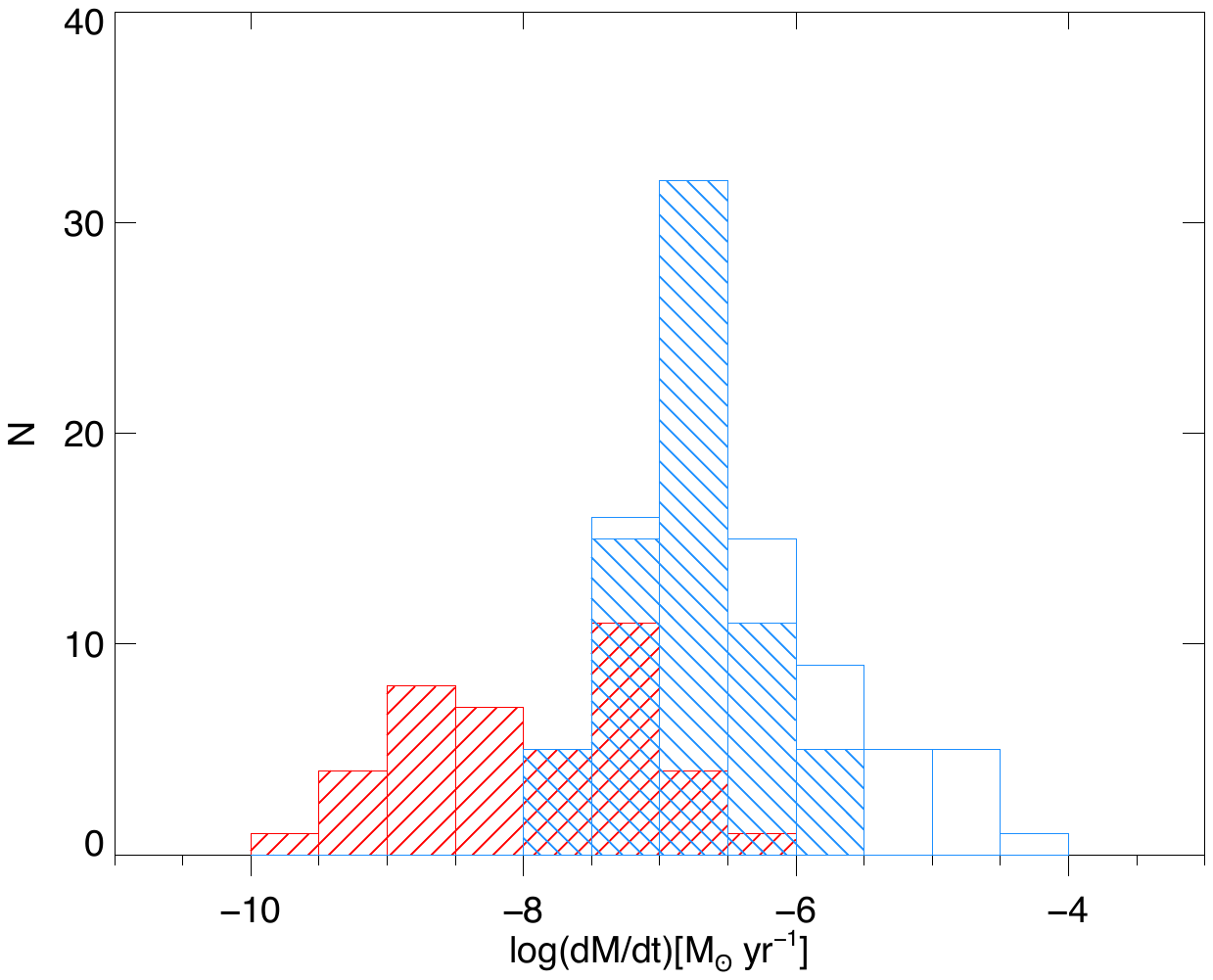}
\caption{Distribution of the stellar accretion rates of all Herbig stars (blue) and IMTTSs (red). The empty bins are Herbig stars with 2.3 $\lesssim$ log(L$_{\star}$/L$_{\sun})$ $\lesssim$2.7.
The median stellar accretion rate of Herbig stars is one order of magnitude higher than the median stellar accretion rate of IMTTSs.}
\label{fig:hist}
\end{center}
\end{figure}

 Herbig stars from \citet{Fairlamb_feb17, Fairlamb_oct15, Vioque_dec18, Wichittanakom_mar20} were chosen to compare their accretion rates with IMTTSs. The Herbig star sample was limited to sources with $\rm 7,200K \leq T_{eff} \leq 13000K$ and $\rm log(L_{\star}) \leq 2.7L_{\sun}$ to select stars with $\rm M_{\star}~\lesssim~4M_{\sun}$ resulting in 89 stars. 
A histogram of the stellar accretion rates of IMTTSs and Herbig stars is presented in Figure~\ref{fig:hist}. The median accretion rate of Herbig stars is $\rm 1.9\times10^{-7}~M_{\sun}~yr^{-1}$), which is over an order of magnitude higher than the median accretion rate of the IMTTSs ($\rm 1.2\times10^{-8}~M_{\sun}~yr^{-1}$). Even if we decrease the log luminosity limit to 2.3L$_{\sun}$ to restrict our sample to stars $\lesssim$ 3.5M$_{\sun}$, the median rate of Herbig stars is $\rm 1.6\times10^{-7}~M_{\sun}~yr^{-1}$ which is still more than an order of magnitude higher than the median rates of IMTTSs. The range of the log of accretion rates for the IMTTSs spans from roughly -10 to -6, while the range seen in Herbig stars spans from roughly -8 to -4. Although sensitivity limits may restrict the identification of Herbig stars accreting at rates less than $\rm 10^{-8}$ M$_{\odot}$ yr$^{-1}$, there is no evidence of a clustering of Herbig stars at this limit. IMTTSs with accretion rates greater than $\rm 10^{-7}$ M$_{\odot}$ yr$^{-1}$ in the volume we consider are unlikely to be missed. In the appendix, we address potential concerns about the reliability of measurements of the stellar accretion rate of Herbig stars and conclude that the inferred accretion rates of Herbig stars published in the literature are reliable.

\begin{figure*}
\begin{center}
\includegraphics[trim = 40mm 10mm 00mm 20mm, clip, width=.75\linewidth, angle=90]{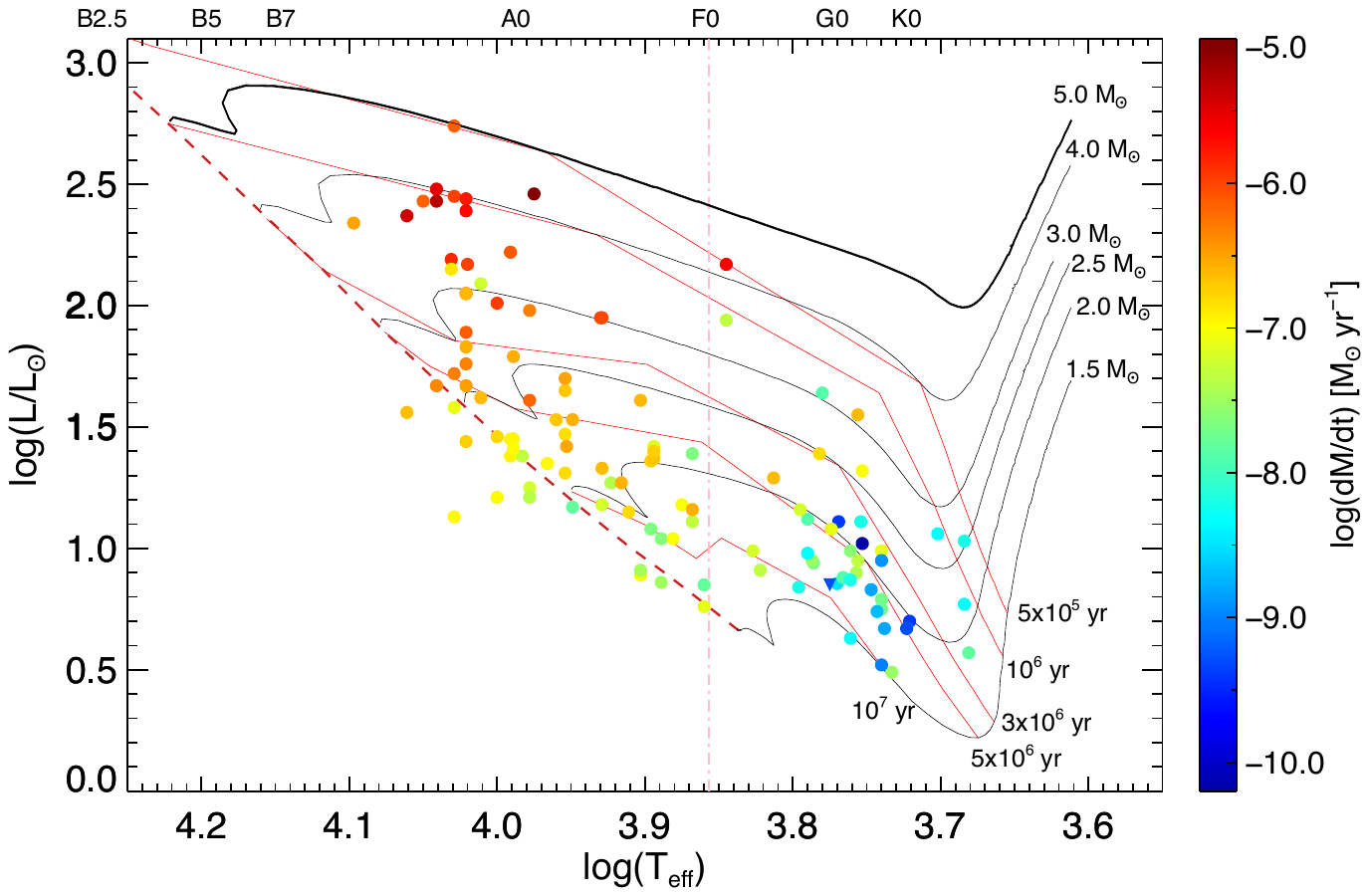}
\caption{Mass accretion rate of IMTTSs and Herbig stars. The stellar mass accretion rate is indicated by the color. The evolutionary tracks and isochrones are from the Siess stellar evolution model \citep{siess_jun00}. The stellar luminosity and temperature were adopted from \citet{Valegard_apr21, Fairlamb_oct15, Vioque_dec18, Wichittanakom_mar20}. The early main sequence is indicated by the red dashed line. The brown dashed line demarcates Herbig stars and IMTTSs. 
\label{fig:hr}}
\end{center}
\end{figure*}

To disentangle the dependency of the stellar accretion rate on stellar mass and age, we plot the mass accretion rates of these IMTTSs and Herbig stars on a Hertzsprung-Russell diagram (Figure \ref{fig:hr}) with isochrones and evolutionary tracks from \citet{siess_jun00}. While there is some inconsistency among the more modern stellar models, the Siess models appear to be the most accurate at the higher mass end of pre-main sequence stars \citep{Braun2021}.  A cursory examination of Figure \ref{fig:hr} immediately reveals that the accretion rates of IMTTSs are systematically lower than the Herbig stars. Tracing the zero-age main sequence (ZAMS), there is a clear trend between the stellar accretion rate and mass/age as noted previously by \citet{Wichittanakom_mar20}. However, if one follows a given mass track (e.g. M$_{\star}$=2M$_{\sun}$) one can see the stellar accretion rate increase from 10$^{-9}$M$_{\sun}$yr$^{-1}$ at about 1-3Myr to 10$^{-7}$M$_{\sun}$yr$^{-1}$ near the ZAMS. 

An increase in the stellar accretion rate with age stands in stark contrast with what is predicted by viscous evolution models with a uniform $\alpha$ where the stellar accretion rate is expected to decline with age \citep{Hartmann_98}. The observed accretion rates of T Tauri stars in different clusters are consistent with such a model \citep[e.g.,][]{Muzerolle_dec04, Sicilia-Aguilar_oct06}. This suggests that the relationship between age and accretion rate for intermediate-mass objects is fundamentally different than that of their lower-mass counterparts.

\section{Modeling Accretion Evolution of Intermediate-Mass Stars}

The relatively high accretion rates of Herbig stars point to a well-known incongruity with the expected mass of the disk \citep[e.g.,][]{mendigutia2012, dong_aug18, Grant_sep23}{}. Indeed, 
the stellar accretion rate of Herbig stars scales weakly with the disk mass inferred from millimeter continuum observations,

\begin{equation}
    {\rm log}(\dot{M}) = (-0.03\pm0.21){\rm log}(M_{disk}) + (-6.99\pm0.52)
\end{equation} 

\noindent for $\rm -4.8\lesssim log(M_{disk}/M_{\sun})\lesssim-1.2$ \citep{Grant_sep23}. The typical Herbig star is a few million years old with a stellar accretion rate of a few $\times$ 10$^{-7}$ M$_{\odot}$ yr$^{-1}$. Estimates of the disk mass inferred from millimeter measurements imply that the lifetime of disks is only about 10\% of the stellar age, assuming the observed stellar accretion rate is constant  \citep{mendigutia2012, dong_aug18}. 
If the accretion rate declines as a power law with  \.{M}(t) $\propto t^{-3/2}$, as expected for viscous accretion with a uniform $\alpha$ \citep{Hartmann_98},  then the minimum mass in the disk can be determined from integrating this expression such that
\begin{equation}
    M_{disk}(t)=2t\dot{M}(t).
    \label{eqn:3}
\end{equation}
As seen in figure \ref{fig:hr}, the typical age and stellar accretion rate of a Herbig disk is a few Myr and 10$^{-7}$ M$_{\odot}$ yr$^{-1}$ respectively.  
If equation \ref{eqn:3} accurately approximates the time evolution of the stellar accretion rate, then the typical disk mass is $\sim$30\% of the Herbig star mass. In this case, the disk mass inferred from millimeter measurements is underestimated by about one order of magnitude, and the typical Herbig disk is gravitationally unstable. Furthermore, the stellar accretion rates would imply implausibly high disk masses during the IMTTS phase. 

Here, we explore a scenario that can account for the increase in the stellar accretion rate as the star evolves toward the main sequence and resolves the apparent incongruity between the inferred stellar accretion rate and disk mass of Herbig stars. 
Unlike their lower-mass counterparts, intermediate-mass stars spend a substantial fraction of their pre-main-sequence evolution along the radiative Henyey tracks. Thus, the temperature of intermediate-mass stars will increase by a factor of 2 or more as they evolve to the ZAMS (Figure~ \ref{fig:hr}). As a result, the luminosity of the far-ultraviolet (FUV) continuum ($\sim$6-13.6~eV) can increase by as much as four orders of magnitude. 

\citet{Kunitomo2021} explored the role of stellar evolution on the photoevaporation of disks around intermediate-mass stars. These authors found that the sharp increase in the FUV luminosity of these stars accelerated the dispersal of the disks. However, the FUV irradiation of disks can also affect how the disks accrete.  

In MHD simulations of wind-driven accretion, the accretion rate depends strongly on the penetration depth of the FUV field and the surface density of the FUV ionized layer in the disk atmosphere \citep{Bai_may13}. Similar conclusions have been found in MHD simulations of accretion flows driven by in-disk angular momentum transport due to the magnetorotational instability (MRI). \citet{Bai_sep11} argued that the total MRI-driven accretion rate in the inner disk is significantly affected by the FUV ionization, and the FUV could even be the main source of ionization in the outer disk. The effects of FUV ionization on surface layer accretion were modeled by \citet{Perez-Becker_jul11} in which the surface density of the MRI active layer scales as $\Sigma_{\star} \propto L^{1/2}_{FUV}$.  

If disks accrete through their surface, and if that accretion (whether driven by disk winds or another in-disk angular momentum transport mechanism) is proportional to the surface density of the ionized layer, then one would expect that the accretion rate would increase as the stellar radiation field hardens. Regardless of the details of how accretion and other mechanisms of disk dispersal (e.g. photoevaporation) affect the disk as the gas is depleted and becomes optically thin, when this occurs, accretion in these systems effectively stops. 

\begin{figure*}[!h]
\begin{center}
\includegraphics[scale=0.7,angle=0, trim= 50mm 30mm 50mm 30mm]{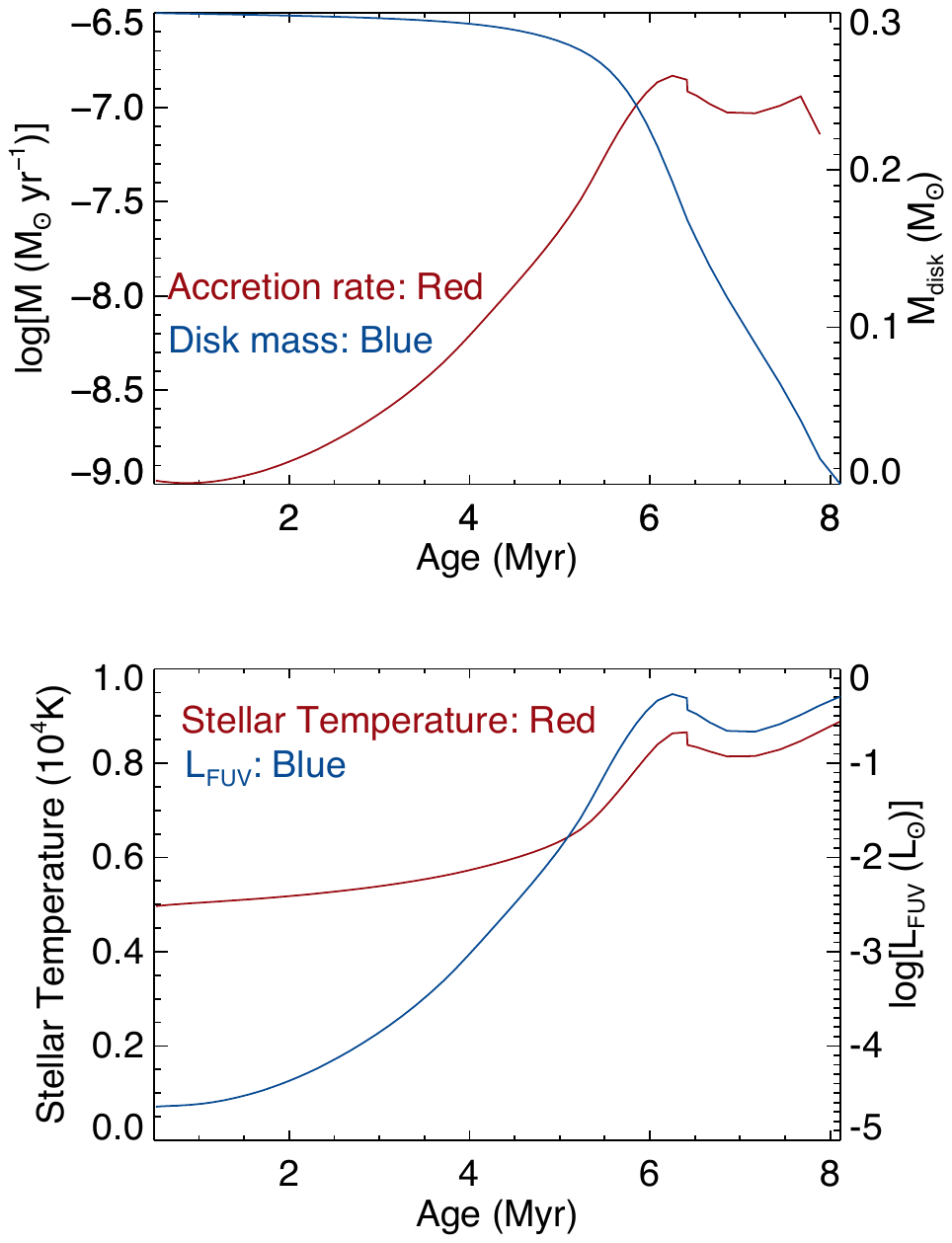}
\caption{Comparison of the accretion rate (red) and disk mass (blue) in the upper panel and the effective temperature (red) and L$_{\rm FUV}$ (blue) in the lower panel during the pre-main-sequence evolution of a 2~M$_{\sun}$ star. The star begins with a 0.15~M$_{\star}$ disk and an accretion rate of 10$^{-9}$M$_{\sun}$yr$^{-1}$. The accretion rate declines as t$^{-3/2}$ until the effect of the FUV radiation of the star dominates. The accretion rate then stays above 10$^{-8}$M$_{\sun}$yr$^{-1}$ for $\sim$4Myr. While accretion is known to be variable and photoevaporation and planet formation may cut off accretion prior to entirely emptying the disk of gas, this scenario qualitatively captures the observed trends among IMTTSs and Herbig stars.  \label{fig:accExample}}
\end{center}
\end{figure*}

To illustrate the effect of stellar evolution on the accretion rate of an intermediate-mass star, we consider a toy model in which the stellar accretion rate initially declines as $t^{-3/2}$ as one might expect for viscous accretion with a constant $\alpha$ \citep[see, for example,][]{Hartmann_98}. 
We add a component to the accretion rate that scales as $\rm L_{FUV}^{1/2}$ \citep{Perez-Becker_jul11}. Thus, the time dependence of the stellar accretion rate is described by

\begin{align}
    \rm \dot{M}(t=&10^{-9}~M_{\sun}~yr^{-1} \left(\frac{t}{\rm 0.5~Myr}\right)^{-\frac{3}{2}}\nonumber \\
     &+2\times10^{-7} ~M_{\sun}~yr^{-1}\left(\frac{L_{FUV}(t)}{ L_{\sun}}\right)^\frac{1}{2}. 
    \label{equation:acc_evol}
\end{align}

We consider a 2M$_{\sun}$ star with a disk that begins just marginally 
gravitationally unstable ($\rm M_{disk}=0.15M_{\star}$) at 0.5Myr (see for example, \citealt{hall2019}). The fiducial stellar accretion rates in equation \ref{equation:acc_evol} were chosen to reflect the typical beginning and ending accretion rates of 2M$_{\sun}$ stars in our sample.
To determine the FUV luminosity, we assume that the FUV is dominated by the continuum emission of the stellar photosphere. While several lines in the FUV scale with the accretion luminosity \citep{Calvet_sep04}, the integrated luminosity of these lines is 1-3 orders of magnitude lower than the continuum luminosity at the stellar accretion rates covered by our sample. We integrate the flux from the stellar models described by \citet{Coelho2014} 
blueward of 2100\AA~for stars ranging from 4500 to 13,000~K. The integrated flux for each time step was determined by interpolating among the measurements from the models. The effective temperature, FUV luminosity, and stellar accretion rate are then calculated for each time step. We assume that the mass of the disk is uniquely determined by the stellar accretion rate (i.e., we ignore the effects of planet formation and photoevaporation for this scenario). 
The stellar accretion rate, disk mass, luminosity of the FUV, and effective temperature of the star are plotted in Figure \ref{fig:accExample}. 

\begin{figure*}[!h]
\begin{center}
\includegraphics[scale=.65,angle=90, trim= 30mm 0mm 0mm 0mm]{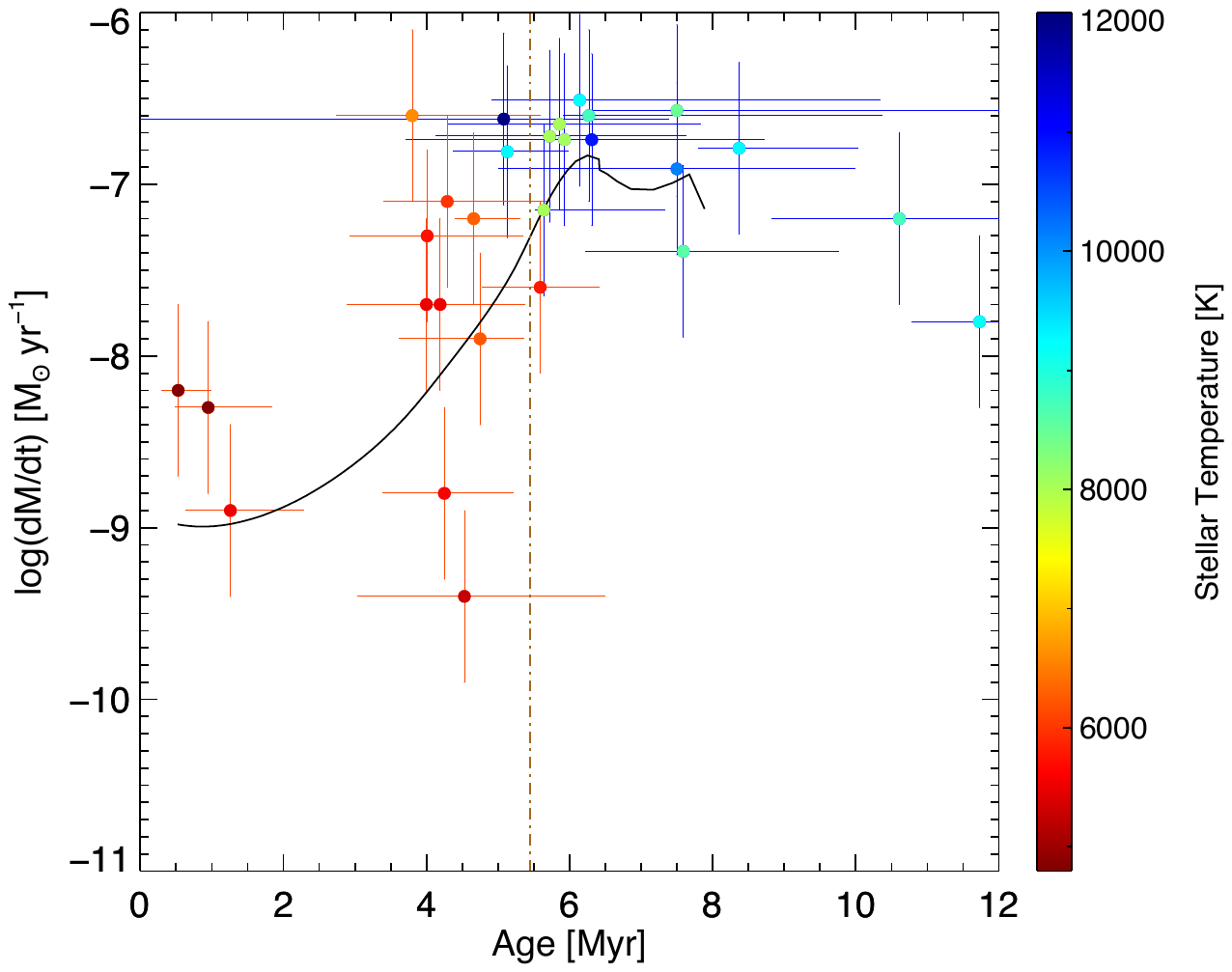}
\caption{The toy model from Figure \ref{fig:accExample} is plotted against sources with masses within 1$\sigma$ of 2~M$_{\sun}$. The IMTTSs are plotted with red error bars, and the Herbig stars are plotted with blue error bars. The color of each data point corresponds to the stellar effective temperature.  \label{fig:model_example}}
\end{center}
\end{figure*}

In this scenario, the disk is fully accreted after 8~Myr, and the Herbig phase ($\rm T_{eff} \gtrsim$ 7,200~K and \.{M}$ \gtrsim 10^{-8}$M$_{\sun}$~yr$^{-1}$) lasts for 2.5~Myr. To compare the toy model with our data, we recalculated the age of our entire sample using the Siess evolutionary tracks. The uncertainty in the age was inferred from the uncertainty in the temperature and luminosity. We adopted a 0.5~dex uncertainty in the accretion rate reflecting the typical uncertainty reported in the literature. Stars with a mass within 1$\sigma$ of 2M$_{\sun}$ were plotted, and the toy model presented in Figure \ref{fig:accExample} was plotted with the data (Figure \ref{fig:model_example}). 

Our toy model is only intended to capture the qualitative trend of the accretion rate as a function of age. Effects such as variable accretion, contribution to the FUV from accretion, photoevaporation, and the effect of a gap-opening planet are not included in this model. However, the scenario we propose qualitatively matches the observed evolution of the stellar accretion rate. The youngest IMTTSs are clustered near an accretion rate of $10^{-9} \rm M_{\sun}~yr^{-1}$. The accretion rate of IMTTSs increases as they approach the Herbig phase and maintain a relatively constant stellar accretion rate near $10^{-7} \rm M_{\sun}~yr^{-1}$. This scenario also accounts for three puzzling trends observed about IMTTSs and Herbig stars. First, \citet{Grant_sep23} note that the accretion rate of Herbig stars is relatively constant over 3~dex in disk mass. This is consistent with what we observe in Figure \ref{fig:accExample} where the stellar accretion rate hovers around $\rm 10^{-7} M_{\sun} yr^{-1}$ as the disk dissipates over 2.5 Myr. This stands in stark contrast to T Tauri stars, where the stellar accretion rate scales with the disk mass nearly linearly \citep{Testi2022}. Second, \citep{Stapper2025} compared the disk mass of IMTTS and Herbig stars. They found that the disk masses were consistent, suggesting that the accretion of disk material onto the star proceeds at a relatively slow rate during the IMTTS phase, as reflected in our model. Third, our model can account for Herbig stars with relatively low disk masses and high accretion rates. 

\section{Discussion and Conclusions}
We have studied the accretion rates of intermediate-mass stars, and find that the median accretion rate of IMTTSs is 1.2 $\times$ 10$^{-8}$ M$_{\odot}$ yr$^{-1}$ - more than one order of magnitude lower than that of their more evolved counterparts, the Herbig stars (1.9 $\times$ 10$^{-7}$ M$_{\odot}$ yr$^{-1}$). The increase in the accretion rate of intermediate-mass stars as they evolve from IMTTSs to Herbig stars can be explained as a consequence of the increase in their FUV luminosity. 

We also showed that an increase in the accretion rate of an intermediate-mass star (M = 2 M$_{\odot}$) which coincides with its pre-main-sequence temperature increase can explain the general trends of these objects seen on the H-R diagram (Figure~\ref{fig:hr}). If the stellar accretion rate scales with the FUV luminosity of the star such that the stellar accretion rate increases as the star evolves, then much lower disk masses are required to sustain the observed accretion rates among Herbig stars than if the stellar accretion rate declines with age. In the scenario we propose, a disk that begins marginally gravitationally unstable will be able to sustain an accretion rate of order 10$^{-7}$M$_{\sun}$yr$^{-1}$ for 2.5~Myr and deplete its disk at an age of 8~Myr (near the ZAMS). 

Insofar as at least some A stars accrete at significantly higher rates than their evolutionary predecessors (i.e., the IMTTSs), this may motivate further investigation into the role of the FUV luminosity of the star in driving accretion. While the toy model we present does not attempt to capture the physical mechanism by which material is transported in the disk, any such model should account for the role of the FUV irradiation of the disk. Here, we discuss two other direct consequences arising from this work.

\subsection{Weak-lined Herbig Stars} 
While some young intermediate-mass stars may have disk masses approaching 15\% of their stellar mass, higher disk masses are unlikely. However, a few Herbig stars with accretion rates of 10$^{-7}$~M$_{\sun}$~yr$^{-1}$ have ages approaching 12~Myr (Figure~\ref{fig:model_example}). To account for this, it is important to note that one of the original selection criteria for Herbig stars was the association with nebulosity \citep{Herbig1960}. There is growing evidence that such nebulae may supply additional material to disks \citep[e.g.,][and references therein]{Gupta2023, Speedie2025}{}. While the replenishment of the disk from surrounding nebulae can account for the relatively long lifetime of some Herbig disks, it is not clear that this process can account for the increase in the accretion rate relative to IMTTSs found here. 

On the other hand, one should not expect all intermediate-mass stars to have a disk mass at 0.5Myr that is 15\% of the stellar mass or that typical intermediate mass stars accrete additional circumstellar material from surrounding nebulosity.  In this case, we might expect a large population of young A stars ($\lesssim5$Myr) that are the intermediate-mass analogs to weak-lined T Tauri stars (i.e., weak-lined Herbig Stars). \citet{Iglesias_mar23} performed a volume-limited survey (d\textless300pc) of intermediate-mass stars  ($1.5M_{\sun} \leq M_{\star} \leq 3.5M_{\sun}$) with an infrared excess. They found 129 nonaccreting objects and estimate that their sample is 35--55\% complete. They classified 112 of these objects as debris disks and 17 of these objects as hybrid disks (i.e., disks in transition from protoplanetary disks to debris disks) following the classification scheme proposed by \citet{Wyatt_jun15}.  The mean age of the debris disks in their sample is 4Myr. In the same volume and mass range, there are 30 Herbig stars, indicating that Herbig stars comprise roughly 10\%  of pre-main-sequence A and B stars. Thus, Herbig stars may reflect the upper end of the initial disk mass distribution captured near the end of their accretion lifetime. As noted in section 4, \citet{Grant_sep23} report the short lifetimes of many disks implied by the disk masses and stellar accretion rates. These short disk lifetimes may reflect the final stages of the transition from the Herbig phase to the “weak-lined” Herbig state.

\subsection{Implications for Spiral Structure as a Signpost of Forming Gas Giant Planets}

The results presented here also bear on the interpretation of the disk morphologies observed among Herbig stars. \citet{dong_aug18} note the ubiquity of Herbig disks that reveal spiral structure relative to their lower mass counterparts. These authors note that this result is consistent with a higher frequency of gas giant planet formation among Herbig stars, but this result could also reflect a higher rate of disks that are not gravitationally stable. The challenge with the latter scenario is the need for long-lived massive disks, which one might naively infer from the high accretion rate of Herbig stars. However, if the scenario we propose is correct, then the stellar accretion rates measured among Herbig stars do not require the high disk masses that are one order of magnitude greater than the values inferred from (sub)millimeter studies \citep{dong_aug18}. Hence, spiral structure among Herbig stars is more plausibly a signpost of gas giant planet formation. Finally, our model allows for disks with sufficient mass to form planets around Herbig stars to persist for several Myr, even at the high accretion rates observed in this evolutionary state, providing time for gas giant planet formation in these systems.

\vspace{5mm}

\noindent This research has made use of the VizieR catalog access tool, CDS,
 Strasbourg, France. The original description 
 of the VizieR service was published in \citet{Ochsenbein2000}. This paper was based on Chapter 4 of the Ph.D. thesis by \citet{kern2023}.
 
\vspace{2mm}

\begin{appendix}

Here we address two potential objections to the results presented here.  One is that
the stellar accretion rate of Herbig stars could be systematically overestimated. 
\citet{dong_aug18} note that a large fraction of Herbig stars are $\rm \lambda$ Bo{\"o}tis 
stars \citep[e.g.,][]{folsom2012, kama2015, Guzman2023} and that these stars show a UV excess in 
the FUV ($\sim$1500\AA\ ) and near-ultraviolet ($\sim$2300\AA\ ; \citealt{gray2017}). One of the 
selection criteria for identifying $\rm \lambda$ Bo{\"{o}}tis candidates is based on Str{\"{o}}mgren photometry, 0\textless$\delta$c$_1$\textless0.3, where c$_1$=(u-v)-(v-b) is a measure of the Balmer discontinuity \citep{Paunzen1997}. Most studies of the stellar accretion rate of Herbig stars are based on the measurement of the Balmer discontinuity, as the accretion shock should veil this feature \citep{Muzerolle_dec04, Fairlamb_oct15}. The Balmer discontinuity can be measured using Johnson photometry where
\begin{equation}
    \Delta D_B=(U-B)_0-(U-B).
\end{equation}
\noindent Transformation of the Johnson colors to Str{\"{o}}mgren colors results in,
\begin{equation}
    \Delta D_B=0.675((u-b)_0-(u-b))\pm0.015,
\end{equation}
\noindent where (u-b)=c$_1$+2m$_1$+2(b-y) \citep{Turner1990}. To investigate the evidence for veiling of the Balmer discontinuity that could lead to a systematic overestimate of the stellar accretion rate, we calculated $\rm \Delta D_B$ from Str{\"{o}}mgren photometry for 41 $\rm \lambda$ Bo{\"{o}}tis stars compiled by \citet{Paunzen1997}. Str{\"{o}}mgren colors for ``normal'' stars were taken from \citet{Dalle2020}. We find that the $\rm < \Delta D_B >$ for the sample is 0.022$\pm$0.015 with a standard deviation of 0.04 with a range of -0.06 \textless $\rm \Delta D_B$ \textless 0.12.  In contrast, $\rm \Delta D_B$ for Herbig stars ranges from 0.05 to 1.05 with the majority of the sources ranging from 0.05 to 0.35 \citep{Fairlamb_oct15}. We conclude that the $\lambda$ Bo{\"{o}}tis phenomenon is unlikely to lead to a systematic overestimate of the accretion rate of Herbig stars. 

A second concern is the assumption that the magnetospheric accretion paradigm that successfully describes stellar accretion for CTTSs also applies to Herbig stars. As noted above, Herbig stars do not tend to have strong, well-ordered magnetic fields, so one might expect the freefall distance to be much closer to the star or even perhaps accrete via a boundary layer. If so, then the accretion rates necessary to produce the observed UV excess would even be higher than inferred from the assumption that the stars are accreting magnetospherically (in the limit of a slowly rotating star, the accretion rate of material passing through a boundary layer would need to be twice that of material falling from infinity to produce the same accretion luminosity; \citealt{Lynden-Bell_sep74}). Furthermore, high-velocity red-shifted emission is observed from these sources \citep{Guimares_oct06}, which is difficult to explain via boundary layer accretion. To calculate the stellar accretion rates for this paper, we assumed that the material was falling from infinity. Given the stronger typical magnetic field strengths of IMTTSs than Herbig stars, it is more likely that the stellar accretion rates of Herbig stars are underestimated relative to IMTTSs. As noted in the main text, we adopted the relationship between the luminosity of H$\alpha$ and the accretion luminosity calibrated against T Tauri stars \citep{Alcala_apr17}. Adopting the calibration for Herbig stars by \citet{Fairlamb_feb17} would increase the lower accretion luminosity sample by as much as a factor of four. While the physical origin of the relationship between $\rm L_{H\alpha}$ and $\rm L_{acc}$ is not well understood, it is likely that the size of the funnel flow from the disk to the star and possibly the relationship between the accretion rate and outflow rate affect this relationship. If so, then the magnetic geometry that determines these relationships is more similar for the convective IMTTSs and CTTSs than it is for Herbig stars. However, calibration of the line luminosity and UV excess among IMTTSs is warranted.

\end{appendix}

\bibliography{ref}{}

@ARTICLE{Gupta2023,
       author = {{Gupta}, A. and {Miotello}, A. and {Manara}, C.~F. and {Williams}, J.~P. and {Facchini}, S. and {Beccari}, G. and {Birnstiel}, T. and {Ginski}, C. and {Hacar}, A. and {K{\"u}ffmeier}, M. and {Testi}, L. and {Tychoniec}, L. and {Yen}, H. -W.},
        title = "{Reflections on nebulae around young stars. A systematic search for late-stage infall of material onto Class II disks}",
      journal = {\aap},
         year = 2023,
        month = feb,
       volume = {670},
          eid = {L8},
        pages = {L8},
          doi = {10.1051/0004-6361/202245254},
archivePrefix = {arXiv},
       eprint = {2301.02994},
 primaryClass = {astro-ph.SR},
       adsurl = {https://ui.adsabs.harvard.edu/abs/2023A&A...670L...8G}
}

@ARTICLE{Kunitomo2021,
       author = {{Kunitomo}, Masanobu and {Ida}, Shigeru and {Takeuchi}, Taku and {Pani{\'c}}, Olja and {Miley}, James M. and {Suzuki}, Takeru K.},
        title = "{Photoevaporative Dispersal of Protoplanetary Disks around Evolving Intermediate-mass Stars}",
      journal = {\apj},
         year = 2021,
        month = mar,
       volume = {909},
       number = {2},
          eid = {109},
        pages = {109},
          doi = {10.3847/1538-4357/abdb2a},
archivePrefix = {arXiv},
       eprint = {2103.07673},
 primaryClass = {astro-ph.EP},
       adsurl = {https://ui.adsabs.harvard.edu/abs/2021ApJ...909..109K}
}

@ARTICLE{Alecian_2013,
       author = {{Alecian}, E. and {Wade}, G.~A. and {Catala}, C. and {Grunhut}, J.~H. and {Landstreet}, J.~D. and {Bagnulo}, S. and {B{\"o}hm}, T. and {Folsom}, C.~P. and {Marsden}, S. and {Waite}, I.},
        title = "{A high-resolution spectropolarimetric survey of Herbig Ae/Be stars - I. Observations and measurements}",
      journal = {\mnras},
         year = 2013,
        month = feb,
       volume = {429},
       number = {2},
        pages = {1001-1026},
          doi = {10.1093/mnras/sts383},
archivePrefix = {arXiv},
       eprint = {1211.2907},
 primaryClass = {astro-ph.SR},
       adsurl = {https://ui.adsabs.harvard.edu/abs/2013MNRAS.429.1001A}
}

@ARTICLE{Alcala_apr17,
       author = {{Alcal{\'a}}, J.~M. and {Manara}, C.~F. and {Natta}, A. and {Frasca}, A. and {Testi}, L. and {Nisini}, B. and {Stelzer}, B. and {Williams}, J.~P. and {Antoniucci}, S. and {Biazzo}, K. and {Covino}, E. and {Esposito}, M. and {Getman}, F. and {Rigliaco}, E.},
        title = "{X-shooter spectroscopy of young stellar objects in Lupus. Accretion properties of class II and transitional objects}",
      journal = {\aap},
         year = 2017,
        month = apr,
       volume = {600},
          eid = {A20},
        pages = {A20},
          doi = {10.1051/0004-6361/201629929},
archivePrefix = {arXiv},
       eprint = {1612.07054},
 primaryClass = {astro-ph.SR},
       adsurl = {https://ui.adsabs.harvard.edu/abs/2017A&A...600A..20A}
}

@ARTICLE{Bai_may13,
       author = {{Bai}, Xue-Ning and {Stone}, James M.},
        title = "{Wind-driven Accretion in Protoplanetary Disks. I. Suppression of the Magnetorotational Instability and Launching of the Magnetocentrifugal Wind}",
      journal = {\apj},
         year = 2013,
        month = may,
       volume = {769},
       number = {1},
          eid = {76},
        pages = {76},
          doi = {10.1088/0004-637X/769/1/76},
archivePrefix = {arXiv},
       eprint = {1301.0318},
 primaryClass = {astro-ph.EP},
       adsurl = {https://ui.adsabs.harvard.edu/abs/2013ApJ...769...76B}
}

@ARTICLE{Brittain_feb23,
       author = {{Brittain}, Sean D. and {Kamp}, Inga and {Meeus}, Gwendolyn and {Oudmaijer}, Ren{\'e} D. and {Waters}, L.~B.~F.~M.},
        title = "{Herbig Stars}",
      journal = {\ssr},
         year = 2023,
        month = feb,
       volume = {219},
       number = {1},
          eid = {7},
        pages = {7},
          doi = {10.1007/s11214-023-00949-z},
archivePrefix = {arXiv},
       eprint = {2301.01165},
 primaryClass = {astro-ph.SR},
       adsurl = {https://ui.adsabs.harvard.edu/abs/2023SSRv..219....7B}
}

@ARTICLE{Guimares_oct06,
   author = {{Guimar{\~a}es}, M.~M. and {Alencar}, S.~H.~P. and {Corradi}, W.~J.~B. and 
	{Vieira}, S.~L.~A.},
    title = "{Stellar parameters and evidence of circumstellar activity for a sample of Herbig Ae/Be stars}",
  journal = {\aap},
     year = 2006,
    month = oct,
   volume = 457,
    pages = {581-589},
      doi = {10.1051/0004-6361:20065005},
   adsurl = {http://adsabs.harvard.edu/abs/2006A%26A...457..581G}
}

@ARTICLE{Lynden-Bell_sep74,
       author = {{Lynden-Bell}, D. and {Pringle}, J.~E.},
        title = "{The evolution of viscous discs and the origin of the nebular variables.}",
      journal = {\mnras},
         year = 1974,
        month = sep,
       volume = {168},
        pages = {603-637},
          doi = {10.1093/mnras/168.3.603},
       adsurl = {https://ui.adsabs.harvard.edu/abs/1974MNRAS.168..603L}
}

@article{Muzerolle_dec04,
doi = {10.1086/425260},
url = {https://dx.doi.org/10.1086/425260},
year = {2004},
month = {dec},
publisher = {},
volume = {617},
number = {1},
pages = {406},
author = {James Muzerolle and Paola D’Alessio and Nuria Calvet and Lee Hartmann},
title = {Magnetospheres and Disk Accretion in Herbig Ae/Be Stars},
journal = {The Astrophysical Journal},
}

@article{Villebrun_19,
	author = {{Villebrun}, F. and {Alecian, E.} and {Hussain, G.} and {Bouvier, J.} and {Folsom, C. P.} and {Lebreton, Y.} and {Amard, L.} and {Charbonnel, C.} and {Gallet, F.} and {Haemmerl\'e, L.} and {B\"ohm, T.} and {Johns-Krull, C.} and {Kochukhov, O.} and {Marsden, S. C.} and {Morin, J.} and {Petit, P.}},
	title = {Magnetic fields of intermediate-mass T Tauri stars - I. Magnetic detections and fundamental stellar parameters},
	DOI= "10.1051/0004-6361/201833545",
	url= "https://doi.org/10.1051/0004-6361/201833545",
	journal = {A\&A},
	year = 2019,
	volume = 622,
	pages = "A72",
}

@ARTICLE{bai_sep11,
       author = {{Bai}, Xue-Ning},
        title = "{Magnetorotational-instability-driven Accretion in Protoplanetary Disks}",
      journal = {\apj},
         year = 2011,
        month = sep,
       volume = {739},
       number = {1},
          eid = {50},
        pages = {50},
          doi = {10.1088/0004-637X/739/1/50},
archivePrefix = {arXiv},
       eprint = {1107.2935},
 primaryClass = {astro-ph.EP},
       adsurl = {https://ui.adsabs.harvard.edu/abs/2011ApJ...739...50B}
}

@ARTICLE{Calvet_sep04,
   author = {{Calvet}, N. and {Muzerolle}, J. and {Brice{\~n}o}, C. and {Hern{\'a}ndez}, J. and 
	{Hartmann}, L. and {Saucedo}, J.~L. and {Gordon}, K.~D.},
    title = "{The Mass Accretion Rates of Intermediate-Mass T Tauri Stars}",
  journal = {\aj},
     year = 2004,
    month = sep,
   volume = 128,
    pages = {1294-1318},
      doi = {10.1086/422733},
   adsurl = {http://adsabs.harvard.edu/abs/2004AJ....128.1294C}
}

@ARTICLE{Donehew_feb11,
   author = {{Donehew}, B. and {Brittain}, S.},
    title = "{Measuring the Stellar Accretion Rates of Herbig Ae/Be Stars}",
  journal = {\aj},
     year = 2011,
    month = feb,
   volume = 141,
      eid = {46},
    pages = {46},
      doi = {10.1088/0004-6256/141/2/46},
   adsurl = {http://adsabs.harvard.edu/abs/2011AJ....141...46D}
}

@Article{dong_aug18,
  author        = {{Dong}, R. and {Najita}, J.~R. and {Brittain}, S.},
  title         = {{Spiral Arms in Disks: Planets or Gravitational Instability?}},
  journal       = {\apj},
  year          = {2018},
  volume        = {862},
  pages         = {103},
  month         = aug,
  adsurl        = {http://adsabs.harvard.edu/abs/2018ApJ...862..103D},
  archiveprefix = {arXiv},
  doi           = {10.3847/1538-4357/aaccfc},
  eid           = {103},
  eprint        = {1806.05183},
  primaryclass  = {astro-ph.SR},
}

@ARTICLE{Fairlamb_oct15,
   author = {{Fairlamb}, J.~R. and {Oudmaijer}, R.~D. and {Mendigut{\'{\i}}a}, I. and 
	{Ilee}, J.~D. and {van den Ancker}, M.~E.},
    title = "{A spectroscopic survey of Herbig Ae/Be stars with X-shooter - I. Stellar parameters and accretion rates}",
  journal = {\mnras},
archivePrefix = "arXiv",
   eprint = {1507.05967},
 primaryClass = "astro-ph.SR",
     year = 2015,
    month = oct,
   volume = 453,
    pages = {976-1001},
      doi = {10.1093/mnras/stv1576},
   adsurl = {http://adsabs.harvard.edu/abs/2015MNRAS.453..976F}
}

@ARTICLE{Fairlamb_feb17,
   author = {{Fairlamb}, J.~R. and {Oudmaijer}, R.~D. and {Mendigutia}, I. and 
	{Ilee}, J.~D. and {van den Ancker}, M.~E.},
    title = "{A spectroscopic survey of Herbig Ae/Be stars with X-Shooter - II. Accretion diagnostic lines}",
  journal = {\mnras},
archivePrefix = "arXiv",
   eprint = {1610.09636},
 primaryClass = "astro-ph.SR",
     year = 2017,
    month = feb,
   volume = 464,
    pages = {4721-4735},
      doi = {10.1093/mnras/stw2643},
   adsurl = {http://adsabs.harvard.edu/abs/2017MNRAS.464.4721F}
}

@ARTICLE{Fang_jul13,
   author = {{Fang}, M. and {Kim}, J.~S. and {van Boekel}, R. and {Sicilia-Aguilar}, A. and 
	{Henning}, T. and {Flaherty}, K.},
    title = "{Young Stellar Objects in Lynds 1641: Disks, Accretion, and Star Formation History}",
  journal = {\apjs},
archivePrefix = "arXiv",
   eprint = {1304.7777},
 primaryClass = "astro-ph.SR",
     year = 2013,
    month = jul,
   volume = 207,
      eid = {5},
    pages = {5},
      doi = {10.1088/0067-0049/207/1/5},
   adsurl = {https://ui.adsabs.harvard.edu/abs/2013ApJS..207....5F}
}

@article{Gregorio-Hetem_02,
author = {Gregorio-Hetem, J. and Hetem Jr, A.},
title = {Classification of a selected sample of weak T Tauri stars},
journal = {Monthly Notices of the Royal Astronomical Society},
volume = {336},
number = {1},
pages = {197-206},
doi = {https://doi.org/10.1046/j.1365-8711.2002.05716.x},
url = {https://onlinelibrary.wiley.com/doi/abs/10.1046/j.1365-8711.2002.05716.x},
eprint = {https://onlinelibrary.wiley.com/doi/pdf/10.1046/j.1365-8711.2002.05716.x},
year = {2002}
}

@ARTICLE{grant_feb22,
       author = {{Grant}, Sierra L. and {Espaillat}, Catherine C. and {Brittain}, Sean and {Scott-Joseph}, Caleb and {Calvet}, Nuria},
        title = "{Tracing Accretion onto Herbig Ae/Be Stars Using the Br{\ensuremath{\gamma}} Line}",
      journal = {\apj},
         year = 2022,
        month = feb,
       volume = {926},
       number = {2},
          eid = {229},
        pages = {229},
          doi = {10.3847/1538-4357/ac450a},
archivePrefix = {arXiv},
       eprint = {2112.10428},
 primaryClass = {astro-ph.SR},
       adsurl = {https://ui.adsabs.harvard.edu/abs/2022ApJ...926..229G}
}

@article{Grant_sep23,
doi = {10.3847/1538-3881/acf128},
url = {https://dx.doi.org/10.3847/1538-3881/acf128},
year = {2023},
month = {sep},
publisher = {The American Astronomical Society},
volume = {166},
number = {4},
pages = {147},
author = {Sierra L. Grant and Lucas M. Stapper and Michiel R. Hogerheijde and Ewine F. van Dishoeck and Sean Brittain and Miguel Vioque},
title = {The \dot{M}–Mdisk Relationship for Herbig Ae/Be Stars: A Lifetime Problem for Disks with Low Masses?},
journal = {The Astronomical Journal},
}

@Article{Hartmann_98,
  author   = {{Hartmann}, L. and {Calvet}, N. and {Gullbring}, E. and {D'Alessio}, P.},
  title    = {{Accretion and the Evolution of T Tauri Disks}},
  journal  = {\apj},
  year     = {1998},
  volume   = {495},
  pages    = {385-400},
  month    = mar,
  adsurl   = {http://adsabs.harvard.edu/abs/1998ApJ...495..385H},
  doi      = {10.1086/305277},
}

@BOOK{Herbig_88,
       author = {{Herbig}, G.~H. and {Bell}, K. Robbin},
        title = "{Third Catalog of Emission-Line Stars of the Orion Population : 3 : 1988}",
         year = 1988,
       adsurl = {https://ui.adsabs.harvard.edu/abs/1988cels.book.....H}
}

@ARTICLE{Iglesias_mar23,
       author = {{Iglesias}, Daniela P. and {Pani{\'c}}, Olja and {van den Ancker}, Mario and {Petr-Gotzens}, Monika G. and {Siess}, Lionel and {Vioque}, Miguel and {Pascucci}, Ilaria and {Oudmaijer}, Ren{\'e} and {Miley}, James},
        title = "{X-shooter survey of young intermediate-mass stars - I. Stellar characterization and disc evolution}",
      journal = {\mnras},
         year = 2023,
        month = mar,
       volume = {519},
       number = {3},
        pages = {3958-3975},
          doi = {10.1093/mnras/stac3619},
archivePrefix = {arXiv},
       eprint = {2212.06791},
 primaryClass = {astro-ph.SR},
       adsurl = {https://ui.adsabs.harvard.edu/abs/2023MNRAS.519.3958I}
}

@article{Janson_feb21,
	doi = {10.1051/0004-6361/202039683},
  
	url = {https://doi.org/10.1051%2F0004-6361%2F202039683},
  
	year = 2021,
	month = {feb},
  
	publisher = {{EDP} Sciences},
  
	volume = {646},
  
	pages = {A164},
  
	author = {Markus Janson and Vito Squicciarini and Philippe Delorme and Raffaele Gratton and Mickaël Bonnefoy and Sabine Reffert and Eric E. Mamajek and Simon C. Eriksson and Arthur Vigan and Maud Langlois and Natalia Engler and Gaël Chauvin and Silvano Desidera and Lucio Mayer and Gabriel-Dominique Marleau and Alexander J. Bohn and Matthias Samland and Michael Meyer and Valentina d'Orazi and Thomas Henning and Sascha Quanz and Matthew Kenworthy and Joseph C. Carson},
  
	title = {{BEAST} begins: sample characteristics and survey performance of the B-star Exoplanet Abundance Study},
  
	journal = {\aap}
}

@ARTICLE{Jensen_sep09,
       author = {{Jensen}, Eric L.~N. and {Cohen}, David H. and {Gagn{\'e}}, Marc},
        title = "{No Transition Disk? Infrared Excess, PAH, H$_{2}$, and X-Rays from the Weak-Lined T Tauri Star DoAr 21}",
      journal = {\apj},
         year = 2009,
        month = sep,
       volume = {703},
       number = {1},
        pages = {252-269},
          doi = {10.1088/0004-637X/703/1/252},
archivePrefix = {arXiv},
       eprint = {0907.4980},
 primaryClass = {astro-ph.SR},
       adsurl = {https://ui.adsabs.harvard.edu/abs/2009ApJ...703..252J}
}

@ARTICLE{Maheswar_apr03,
       author = {{Maheswar}, G. and {Manoj}, P. and {Bhatt}, H.~C.},
        title = "{VizieR Online Data Catalog: Stephenson H{\ensuremath{\alpha}} stars (Maheswar+, 2003)}",
      journal = {VizieR Online Data Catalog},
         year = 2003,
        month = apr,
          eid = {J/A+A/402/963},
        pages = {J/A+A/402/963},
          doi = {10.26093/cds/vizier.34020963},
       adsurl = {https://ui.adsabs.harvard.edu/abs/2003yCat..34020963M}
}

@article{Manara_14,
	author = {{Manara}, C. F. and {Testi, L.} and {Natta, A.} and {Rosotti, G.} and {Benisty, M.} and {Ercolano, B.} and {Ricci, L.}},
	title = {Gas content of transitional disks: a VLT/X-Shooter study of
          accretion and winds},
	DOI= "10.1051/0004-6361/201323318",
	url= "https://doi.org/10.1051/0004-6361/201323318",
	journal = {A\&A},
	year = 2014,
	volume = 568,
	pages = "A18",
	month = "",
}

@ARTICLE{Manara_aug17,
       author = {{Manara}, C.~F. and {Testi}, L. and {Herczeg}, G.~J. and {Pascucci}, I. and {Alcal{\'a}}, J.~M. and {Natta}, A. and {Antoniucci}, S. and {Fedele}, D. and {Mulders}, G.~D. and {Henning}, T. and {Mohanty}, S. and {Prusti}, T. and {Rigliaco}, E.},
        title = "{X-shooter study of accretion in Chamaeleon I. II. A steeper increase of accretion with stellar mass for very low-mass stars?}",
      journal = {\aap},
         year = 2017,
        month = aug,
       volume = {604},
          eid = {A127},
        pages = {A127},
          doi = {10.1051/0004-6361/201630147},
archivePrefix = {arXiv},
       eprint = {1704.02842},
 primaryClass = {astro-ph.SR},
       adsurl = {https://ui.adsabs.harvard.edu/abs/2017A&A...604A.127M}
}

@ARTICLE{Natta_jun06,
       author = {{Natta}, A. and {Testi}, L. and {Randich}, S.},
        title = "{Accretion in the {\ensuremath{\rho}}-Ophiuchi pre-main sequence stars}",
      journal = {\aap},
         year = 2006,
        month = jun,
       volume = {452},
       number = {1},
        pages = {245-252},
          doi = {10.1051/0004-6361:20054706},
archivePrefix = {arXiv},
       eprint = {astro-ph/0602618},
 primaryClass = {astro-ph},
       adsurl = {https://ui.adsabs.harvard.edu/abs/2006A&A...452..245N}
}

@ARTICLE{Perez-Becker_jul11,
       author = {{Perez-Becker}, Daniel and {Chiang}, Eugene},
        title = "{Surface Layer Accretion in Conventional and Transitional Disks Driven by Far-ultraviolet Ionization}",
      journal = {\apj},
         year = 2011,
        month = jul,
       volume = {735},
       number = {1},
          eid = {8},
        pages = {8},
          doi = {10.1088/0004-637X/735/1/8},
archivePrefix = {arXiv},
       eprint = {1104.2320},
 primaryClass = {astro-ph.EP},
       adsurl = {https://ui.adsabs.harvard.edu/abs/2011ApJ...735....8P}
}

@ARTICLE{Rojas_jul08,
       author = {{Rojas}, G. and {Gregorio-Hetem}, J. and {Hetem}, A.},
        title = "{Towards the main sequence: detailed analysis of weak line and post-T Tauri stars}",
      journal = {\mnras},
         year = 2008,
        month = jul,
       volume = {387},
       number = {3},
        pages = {1335-1343},
          doi = {10.1111/j.1365-2966.2008.13355.x},
       adsurl = {https://ui.adsabs.harvard.edu/abs/2008MNRAS.387.1335R}
}

@article{Sicilia-Aguilar_oct06,
doi = {10.1086/508058},
url = {https://dx.doi.org/10.1086/508058},
year = {2006},
month = {oct},
publisher = {},
volume = {132},
number = {5},
pages = {2135},
author = {Aurora Sicilia-Aguilar and Lee W. Hartmann and Gábor Fürész and Thomas Henning and Cornelis Dullemond and Wolfgang Brandner},
title = {High-Resolution Spectroscopy in Tr 37: Gas Accretion Evolution in Evolved Dusty Disks*},
journal = {The Astronomical Journal}
}

@ARTICLE{siess_jun00,
   author = {{Siess}, L. and {Dufour}, E. and {Forestini}, M.},
    title = "{An internet server for pre-main sequence tracks of low- and intermediate-mass stars}",
  journal = {\aap},
   eprint = {astro-ph/0003477},
     year = 2000,
    month = jun,
   volume = 358,
    pages = {593-599},
   adsurl = {http://adsabs.harvard.edu/abs/2000A%26A...358..593S}
}

@ARTICLE{Simon_nov16,
       author = {{Simon}, M.~N. and {Pascucci}, I. and {Edwards}, S. and {Feng}, W. and {Gorti}, U. and {Hollenbach}, D. and {Rigliaco}, E. and {Keane}, J.~T.},
        title = "{Tracing Slow Winds from T Tauri Stars via Low-velocity Forbidden Line Emission}",
      journal = {\apj},
         year = 2016,
        month = nov,
       volume = {831},
       number = {2},
          eid = {169},
        pages = {169},
          doi = {10.3847/0004-637X/831/2/169},
archivePrefix = {arXiv},
       eprint = {1608.06992},
 primaryClass = {astro-ph.SR},
       adsurl = {https://ui.adsabs.harvard.edu/abs/2016ApJ...831..169S}
}

@article{Stapper_feb22,
	doi = {10.1051/0004-6361/202142164},
  
	url = {https://doi.org/10.1051%2F0004-6361%2F202142164},
  
	year = 2022,
	month = {feb},
  
	publisher = {{EDP} Sciences},
  
	volume = {658},
  
	pages = {A112},
  
	author = {L. M. Stapper and M. R. Hogerheijde and E. F. van Dishoeck and R. Mentel},
  
	title = {The mass and size of Herbig disks as seen by {ALMA}
},
  
	journal = {\aap}
}

@ARTICLE{Valegard_apr21,
       author = {{Valeg{\r{a}}rd}, P. -G. and {Waters}, L.~B.~F.~M. and {Dominik}, C.},
        title = "{What happened before?. Disks around the precursors of young Herbig Ae/Be stars}",
      journal = {\aap},
         year = 2021,
        month = aug,
       volume = {652},
          eid = {A133},
        pages = {A133},
          doi = {10.1051/0004-6361/202039802},
archivePrefix = {arXiv},
       eprint = {2104.14212},
 primaryClass = {astro-ph.SR},
       adsurl = {https://ui.adsabs.harvard.edu/abs/2021A&A...652A.133V}
}

@ARTICLE{Vioque_dec18,
       author = {{Vioque}, M. and {Oudmaijer}, R.~D. and {Baines}, D. and {Mendigut{\'\i}a}, I. and {P{\'e}rez-Mart{\'\i}nez}, R.},
        title = "{Gaia DR2 study of Herbig Ae/Be stars}",
      journal = {\aap},
         year = 2018,
        month = dec,
       volume = {620},
          eid = {A128},
        pages = {A128},
          doi = {10.1051/0004-6361/201832870},
archivePrefix = {arXiv},
       eprint = {1808.00476},
 primaryClass = {astro-ph.SR},
       adsurl = {https://ui.adsabs.harvard.edu/abs/2018A&A...620A.128V}
}

@ARTICLE{Walter_aug92,
       author = {{Walter}, Frederick M.},
        title = "{X-Ray Sources in Regions of Star Formation. IV. High-Resolution Optical Spectroscopy of X-Ray Identified Candidate Pre-Main-Sequence Stars in the Chamaeleon I Cloud}",
      journal = {\aj},
         year = 1992,
        month = aug,
       volume = {104},
        pages = {758},
          doi = {10.1086/116271},
       adsurl = {https://ui.adsabs.harvard.edu/abs/1992AJ....104..758W}
}

@ARTICLE{Wichittanakom_mar20,
       author = {{Wichittanakom}, C. and {Oudmaijer}, R.~D. and {Fairlamb}, J.~R. and
         {Mendigut{\'\i}a}, I. and {Vioque}, M. and {Ababakr}, K.~M.},
        title = "{The accretion rates and mechanisms of Herbig Ae/Be stars}",
      journal = {\mnras},
         year = 2020,
        month = mar,
       volume = {493},
       number = {1},
        pages = {234-249},
          doi = {10.1093/mnras/staa169},
archivePrefix = {arXiv},
       eprint = {2001.05971},
 primaryClass = {astro-ph.SR},
       adsurl = {https://ui.adsabs.harvard.edu/abs/2020MNRAS.493..234W}
}

@ARTICLE{Wyatt_jun15,
       author = {{Wyatt}, M.~C. and {Pani{\'c}}, O. and {Kennedy}, G.~M. and {Matr{\`a}}, L.},
        title = "{Five steps in the evolution from protoplanetary to debris disk}",
      journal = {\apss},
         year = 2015,
        month = jun,
       volume = {357},
       number = {2},
          eid = {103},
        pages = {103},
          doi = {10.1007/s10509-015-2315-6},
archivePrefix = {arXiv},
       eprint = {1412.5598},
 primaryClass = {astro-ph.EP},
       adsurl = {https://ui.adsabs.harvard.edu/abs/2015Ap&SS.357..103W}
}

@ARTICLE{mendigutia2020,
       author = {{Mendigut{\'\i}a}, Ignacio},
        title = "{On the Mass Accretion Rates of Herbig Ae/Be Stars. Magnetospheric Accretion or Boundary Layer?}",
      journal = {Galaxies},
         year = 2020,
        month = may,
       volume = {8},
       number = {2},
        pages = {39},
          doi = {10.3390/galaxies8020039},
archivePrefix = {arXiv},
       eprint = {2005.01745},
 primaryClass = {astro-ph.SR},
       adsurl = {https://ui.adsabs.harvard.edu/abs/2020Galax...8...39M}
}

@ARTICLE{JohnsKrull2007,
       author = {{Johns-Krull}, Christopher M.},
        title = "{The Magnetic Fields of Classical T Tauri Stars}",
      journal = {\apj},
         year = 2007,
        month = aug,
       volume = {664},
       number = {2},
        pages = {975-985},
          doi = {10.1086/519017},
archivePrefix = {arXiv},
       eprint = {0704.2923},
 primaryClass = {astro-ph},
       adsurl = {https://ui.adsabs.harvard.edu/abs/2007ApJ...664..975J}
}

@ARTICLE{Vinkmodel2005,
       author = {{Vink}, Jorick S. and {Harries}, T.~J. and {Drew}, J.~E.},
        title = "{Polarimetric line profiles for scattering off rotating disks}",
      journal = {\aap},
         year = 2005,
        month = jan,
       volume = {430},
        pages = {213-222},
          doi = {10.1051/0004-6361:20041463},
archivePrefix = {arXiv},
       eprint = {astro-ph/0409512},
 primaryClass = {astro-ph},
       adsurl = {https://ui.adsabs.harvard.edu/abs/2005A&A...430..213V}
}

@ARTICLE{Vink2005,
   author = {{Vink}, J.~S. and {Drew}, J.~E. and {Harries}, T.~J. and {Oudmaijer}, R.~D. and 
	{Unruh}, Y.},
    title = "{Probing the circumstellar structures of T Tauri stars and their relationship to those of Herbig stars}",
  journal = {\mnras},
   eprint = {astro-ph/0502535},
     year = 2005,
    month = may,
   volume = 359,
    pages = {1049-1064},
      doi = {10.1111/j.1365-2966.2005.08969.x},
   adsurl = {http://adsabs.harvard.edu/abs/2005MNRAS.359.1049V}
}

@ARTICLE{Vink2002,
   author = {{Vink}, J.~S. and {Drew}, J.~E. and {Harries}, T.~J. and {Oudmaijer}, R.~D.
	},
    title = "{Probing the circumstellar structure of Herbig Ae/Be stars}",
  journal = {\mnras},
   eprint = {astro-ph/0208137},
     year = 2002,
    month = nov,
   volume = 337,
    pages = {356-368},
      doi = {10.1046/j.1365-8711.2002.05920.x},
   adsurl = {https://ui.adsabs.harvard.edu/abs/2002MNRAS.337..356V}
}

@ARTICLE{Ababakr2017,
   author = {{Ababakr}, K.~M. and {Oudmaijer}, R.~D. and {Vink}, J.~S.},
    title = "{A statistical spectropolarimetric study of Herbig Ae/Be stars}",
  journal = {\mnras},
archivePrefix = "arXiv",
   eprint = {1707.08408},
 primaryClass = "astro-ph.SR",
     year = 2017,
    month = nov,
   volume = 472,
    pages = {854-868},
      doi = {10.1093/mnras/stx1891},
   adsurl = {http://adsabs.harvard.edu/abs/2017MNRAS.472..854A}
}

@ARTICLE{Ababakr2016,
       author = {{Ababakr}, K.~M. and {Oudmaijer}, R.~D. and {Vink}, J.~S.},
        title = "{Linear spectropolarimetry across the optical spectrum of Herbig Ae/Be stars}",
      journal = {\mnras},
         year = 2016,
        month = sep,
       volume = {461},
       number = {3},
        pages = {3089-3110},
          doi = {10.1093/mnras/stw1534},
archivePrefix = {arXiv},
       eprint = {1607.02440},
 primaryClass = {astro-ph.SR},
       adsurl = {https://ui.adsabs.harvard.edu/abs/2016MNRAS.461.3089A}
}

@ARTICLE{Vink2015,
       author = {{Vink}, Jorick S.},
        title = "{Linear line spectropolarimetry of Herbig Ae/Be stars}",
      journal = {\apss},
         year = 2015,
        month = jun,
       volume = {357},
       number = {2},
          eid = {98},
        pages = {98},
          doi = {10.1007/s10509-015-2323-6},
archivePrefix = {arXiv},
       eprint = {1501.07436},
 primaryClass = {astro-ph.SR},
       adsurl = {https://ui.adsabs.harvard.edu/abs/2015Ap&SS.357...98V}
}

@ARTICLE{CCM1989,
       author = {{Cardelli}, Jason A. and {Clayton}, Geoffrey C. and {Mathis}, John S.},
        title = "{The Relationship between Infrared, Optical, and Ultraviolet Extinction}",
      journal = {\apj},
         year = 1989,
        month = oct,
       volume = {345},
        pages = {245},
          doi = {10.1086/167900},
       adsurl = {https://ui.adsabs.harvard.edu/abs/1989ApJ...345..245C}
}

@ARTICLE{Coelho2014,
       author = {{Coelho}, P.~R.~T.},
        title = "{A new library of theoretical stellar spectra with scaled-solar and {\ensuremath{\alpha}}-enhanced mixtures}",
      journal = {\mnras},
         year = 2014,
        month = may,
       volume = {440},
       number = {2},
        pages = {1027-1043},
          doi = {10.1093/mnras/stu365},
archivePrefix = {arXiv},
       eprint = {1404.3243},
 primaryClass = {astro-ph.SR},
       adsurl = {https://ui.adsabs.harvard.edu/abs/2014MNRAS.440.1027C}
}

@ARTICLE{mendigutia2012,
       author = {{Mendigut{\'\i}a}, I. and {Mora}, A. and {Montesinos}, B. and {Eiroa}, C. and {Meeus}, G. and {Mer{\'\i}n}, B. and {Oudmaijer}, R.~D.},
        title = "{Accretion-related properties of Herbig Ae/Be stars. Comparison with T Tauris}",
      journal = {\aap},
         year = 2012,
        month = jul,
       volume = {543},
          eid = {A59},
        pages = {A59},
          doi = {10.1051/0004-6361/201219110},
archivePrefix = {arXiv},
       eprint = {1205.4734},
 primaryClass = {astro-ph.SR},
       adsurl = {https://ui.adsabs.harvard.edu/abs/2012A&A...543A..59M}
}

@ARTICLE{folsom2012,
       author = {{Folsom}, C.~P. and {Bagnulo}, S. and {Wade}, G.~A. and {Alecian}, E. and {Landstreet}, J.~D. and {Marsden}, S.~C. and {Waite}, I.~A.},
        title = "{Chemical abundances of magnetic and non-magnetic Herbig Ae/Be stars}",
      journal = {\mnras},
         year = 2012,
        month = may,
       volume = {422},
       number = {3},
        pages = {2072-2101},
          doi = {10.1111/j.1365-2966.2012.20718.x},
archivePrefix = {arXiv},
       eprint = {1202.1845},
 primaryClass = {astro-ph.SR},
       adsurl = {https://ui.adsabs.harvard.edu/abs/2012MNRAS.422.2072F}
}

@ARTICLE{kama2015,
       author = {{Kama}, M. and {Folsom}, C.~P. and {Pinilla}, P.},
        title = "{Fingerprints of giant planets in the photospheres of Herbig stars}",
      journal = {\aap},
         year = 2015,
        month = oct,
       volume = {582},
          eid = {L10},
        pages = {L10},
          doi = {10.1051/0004-6361/201527094},
archivePrefix = {arXiv},
       eprint = {1509.02741},
 primaryClass = {astro-ph.SR},
       adsurl = {https://ui.adsabs.harvard.edu/abs/2015A&A...582L..10K}
}

@ARTICLE{Gray2017,
       author = {{Gray}, R.~O. and {Riggs}, Q.~S. and {Koen}, C. and {Murphy}, S.~J. and {Newsome}, I.~M. and {Corbally}, C.~J. and {Cheng}, K. -P. and {Neff}, J.~E.},
        title = "{The Discovery of {\ensuremath{\lambda}} Bootis Stars: The Southern Survey I}",
      journal = {\aj},
         year = 2017,
        month = jul,
       volume = {154},
       number = {1},
          eid = {31},
        pages = {31},
          doi = {10.3847/1538-3881/aa6d5e},
       adsurl = {https://ui.adsabs.harvard.edu/abs/2017AJ....154...31G}
}

@ARTICLE{Turner1990,
       author = {{Turner}, David G.},
        title = "{Transformations between Stromgren and UBV Colors for Early-Type Stars}",
      journal = {\pasp},
         year = 1990,
        month = nov,
       volume = {102},
        pages = {1331},
          doi = {10.1086/132769},
       adsurl = {https://ui.adsabs.harvard.edu/abs/1990PASP..102.1331T}
}

@ARTICLE{Paunzen1997,
       author = {{Paunzen}, E. and {Weiss}, W.~W. and {Heiter}, U. and {North}, P.},
        title = "{A consolidated catalogue of lambda Bootis stars}",
      journal = {\aaps},
         year = 1997,
        month = may,
       volume = {123},
        pages = {93-101},
          doi = {10.1051/aas:1997308},
       adsurl = {https://ui.adsabs.harvard.edu/abs/1997A&AS..123...93P}
}

@ARTICLE{Dalle2020,
       author = {{Dalle Mese}, G. and {L{\'o}pez-Cruz}, O. and {Schuster}, W.~J. and {Chavarr{\'\i}a-K}, C. and {Ibarra-Medel}, H.~J.},
        title = "{The average physical properties of A-G stars derived from uvby-H {\ensuremath{\beta}} Str{\"o}mgren-Crawford photometry as the basis for a spectral-classification synthetical approach}",
      journal = {\mnras},
         year = 2020,
        month = may,
       volume = {494},
       number = {2},
        pages = {2995-3013},
          doi = {10.1093/mnras/staa816},
archivePrefix = {arXiv},
       eprint = {2003.09563},
 primaryClass = {astro-ph.SR},
       adsurl = {https://ui.adsabs.harvard.edu/abs/2020MNRAS.494.2995D}
}

@ARTICLE{mendigutia2011,
       author = {{Mendigut{\'\i}a}, I. and {Calvet}, N. and {Montesinos}, B. and {Mora}, A. and {Muzerolle}, J. and {Eiroa}, C. and {Oudmaijer}, R.~D. and {Mer{\'\i}n}, B.},
        title = "{Accretion rates and accretion tracers of Herbig Ae/Be stars}",
      journal = {\aap},
         year = 2011,
        month = nov,
       volume = {535},
          eid = {A99},
        pages = {A99},
          doi = {10.1051/0004-6361/201117444},
archivePrefix = {arXiv},
       eprint = {1109.3288},
 primaryClass = {astro-ph.SR},
       adsurl = {https://ui.adsabs.harvard.edu/abs/2011A&A...535A..99M}
}

@ARTICLE{Eisner2007,
       author = {{Eisner}, J.~A. and {Hillenbrand}, L.~A. and {White}, R.~J. and {Bloom}, J.~S. and {Akeson}, R.~L. and {Blake}, C.~H.},
        title = "{Near-Infrared Interferometric, Spectroscopic, and Photometric Monitoring of T Tauri Inner Disks}",
      journal = {\apj},
         year = 2007,
        month = nov,
       volume = {669},
       number = {2},
        pages = {1072-1084},
          doi = {10.1086/521874},
archivePrefix = {arXiv},
       eprint = {0707.3833},
 primaryClass = {astro-ph},
       adsurl = {https://ui.adsabs.harvard.edu/abs/2007ApJ...669.1072E}
}

@ARTICLE{Flaherty2008,
       author = {{Flaherty}, K.~M. and {Muzerolle}, J.},
        title = "{Evidence for Early Circumstellar Disk Evolution in NGC 2068/71}",
      journal = {\aj},
         year = 2008,
        month = mar,
       volume = {135},
       number = {3},
        pages = {966-983},
          doi = {10.1088/0004-6256/135/3/966},
archivePrefix = {arXiv},
       eprint = {0712.1601},
 primaryClass = {astro-ph},
       adsurl = {https://ui.adsabs.harvard.edu/abs/2008AJ....135..966F}
}

@ARTICLE{hall2019,
       author = {{Hall}, Cassandra and {Dong}, Ruobing and {Rice}, Ken and {Harries}, Tim J. and {Najita}, Joan and {Alexander}, Richard and {Brittain}, Sean},
        title = "{The Temporal Requirements of Directly Observing Self-gravitating Spiral Waves in Protoplanetary Disks with ALMA}",
      journal = {\apj},
         year = 2019,
        month = feb,
       volume = {871},
       number = {2},
          eid = {228},
        pages = {228},
          doi = {10.3847/1538-4357/aafac2},
archivePrefix = {arXiv},
       eprint = {1901.02407},
 primaryClass = {astro-ph.EP},
       adsurl = {https://ui.adsabs.harvard.edu/abs/2019ApJ...871..228H}
}

@ARTICLE{Herbig1960,
       author = {{Herbig}, George H.},
        title = "{The Spectra of Be- and Ae-Type Stars Associated with Nebulosity}",
      journal = {\apjs},
         year = 1960,
        month = mar,
       volume = {4},
        pages = {337},
          doi = {10.1086/190050},
       adsurl = {https://ui.adsabs.harvard.edu/abs/1960ApJS....4..337H}
}

@ARTICLE{Najita2015,
       author = {{Najita}, Joan R. and {Andrews}, Sean M. and {Muzerolle}, James},
        title = "{Demographics of transition discs in Ophiuchus and Taurus}",
      journal = {\mnras},
         year = 2015,
        month = jul,
       volume = {450},
       number = {4},
        pages = {3559-3567},
          doi = {10.1093/mnras/stv839},
archivePrefix = {arXiv},
       eprint = {1504.05198},
 primaryClass = {astro-ph.SR},
       adsurl = {https://ui.adsabs.harvard.edu/abs/2015MNRAS.450.3559N}
}

@ARTICLE{Guzman2023,
       author = {{Guzm{\'a}n-D{\'\i}az}, J. and {Montesinos}, B. and {Mendigut{\'\i}a}, I. and {Kama}, M. and {Meeus}, G. and {Vioque}, M. and {Oudmaijer}, R.~D. and {Villaver}, E.},
        title = "{Relation between metallicities and spectral energy distributions of Herbig Ae/Be stars. A potential link with planet formation}",
      journal = {\aap},
         year = 2023,
        month = mar,
       volume = {671},
          eid = {A140},
        pages = {A140},
          doi = {10.1051/0004-6361/202245427},
archivePrefix = {arXiv},
       eprint = {2212.14022},
 primaryClass = {astro-ph.SR},
       adsurl = {https://ui.adsabs.harvard.edu/abs/2023A&A...671A.140G}
}

@ARTICLE{Stapper2025,
       author = {{Stapper}, L.~M. and {Hogerheijde}, M.~R. and {van Dishoeck}, E.~F. and {Vioque}, M. and {Williams}, J.~P. and {Ginski}, C.},
        title = "{Intermediate mass T Tauri disk masses and a comparison to their Herbig disk descendants}",
      journal = {\aap},
         year = 2025,
        month = jan,
       volume = {693},
          eid = {A286},
        pages = {A286},
          doi = {10.1051/0004-6361/202450260},
archivePrefix = {arXiv},
       eprint = {2411.08953},
 primaryClass = {astro-ph.EP},
       adsurl = {https://ui.adsabs.harvard.edu/abs/2025A&A...693A.286S}
}

@ARTICLE{Speedie2025,
       author = {{Speedie}, Jessica and {Dong}, Ruobing and {Teague}, Richard and {Segura-Cox}, Dominique and {Pineda}, Jaime E. and {Calcino}, Josh and {Longarini}, Cristiano and {Hall}, Cassandra and {Tang}, Ya-Wen and {Hashimoto}, Jun and {Paneque-Carre{\~n}o}, Teresa and {Lodato}, Giuseppe and {Veronesi}, Bennedetta},
        title = "{Mapping the Merging Zone of Late Infall in the AB Aur Planet-forming System}",
      journal = {\apjl},
         year = 2025,
        month = mar,
       volume = {981},
       number = {2},
          eid = {L30},
        pages = {L30},
          doi = {10.3847/2041-8213/adb7d5},
archivePrefix = {arXiv},
       eprint = {2503.01957},
 primaryClass = {astro-ph.EP},
       adsurl = {https://ui.adsabs.harvard.edu/abs/2025ApJ...981L..30S}
}

@ARTICLE{Testi2022,
       author = {{Testi}, L. and {Natta}, A. and {Manara}, C.~F. and {de Gregorio Monsalvo}, I. and {Lodato}, G. and {Lopez}, C. and {Muzic}, K. and {Pascucci}, I. and {Sanchis}, E. and {Miranda}, A. Santamaria and {Scholz}, A. and {De Simone}, M. and {Williams}, J.~P.},
        title = "{The protoplanetary disk population in the {\ensuremath{\rho}}-Ophiuchi region L1688 and the time evolution of Class II YSOs}",
      journal = {\aap},
         year = 2022,
        month = jul,
       volume = {663},
          eid = {A98},
        pages = {A98},
          doi = {10.1051/0004-6361/202141380},
archivePrefix = {arXiv},
       eprint = {2201.04079},
 primaryClass = {astro-ph.SR},
       adsurl = {https://ui.adsabs.harvard.edu/abs/2022A&A...663A..98T}
}

@ARTICLE{Braun2021,
       author = {{Braun}, Teresa A.~M. and {Yen}, Hsi-Wei and {Koch}, Patrick M. and {Manara}, Carlo F. and {Miotello}, Anna and {Testi}, Leonardo},
        title = "{Dynamical Stellar Masses of Pre-main-sequence Stars in Lupus and Taurus Obtained with ALMA Surveys in Comparison with Stellar Evolutionary Models}",
      journal = {\apj},
         year = 2021,
        month = feb,
       volume = {908},
       number = {1},
          eid = {46},
        pages = {46},
          doi = {10.3847/1538-4357/abd24f},
archivePrefix = {arXiv},
       eprint = {2012.07441},
 primaryClass = {astro-ph.SR},
       adsurl = {https://ui.adsabs.harvard.edu/abs/2021ApJ...908...46B}
}

@ARTICLE{reipurth1996,
       author = {{Reipurth}, B. and {Pedrosa}, A. and {Lago}, M.~T.~V.~T.},
        title = "{H{\ensuremath{\alpha}} emission in pre-main sequence stars. I. an atlas of line profiles.}",
      journal = {\aaps},
         year = 1996,
        month = dec,
       volume = {120},
        pages = {229-256},
       adsurl = {https://ui.adsabs.harvard.edu/abs/1996A&AS..120..229R}
}

@phdthesis{kern2023,
author={Kern,Joshua},
year={2023},
title={An Investigation of the Accretion Processes in T Tauri and Herbig Ae/Be Systems Using High Resolution Optical and Near-Infrared Spectroscopy},
journal={ProQuest Dissertations and Theses},
pages={147},
note={Copyright - Database copyright ProQuest LLC; ProQuest does not claim copyright in the individual underlying works; Last updated - 2024-11-04},
school = {Clemson University, South Carolina},
isbn={9798382649603},
language={English},
url={https://www.proquest.com/dissertations-theses/investigation-accretion-processes-t-tauri-herbig/docview/3059440062/se-2},
}

@ARTICLE{Ochsenbein2000,
       author = {{Ochsenbein}, F. and {Bauer}, P. and {Marcout}, J.},
        title = "{The VizieR database of astronomical catalogues}",
      journal = {\aaps},
         year = 2000,
        month = apr,
       volume = {143},
        pages = {23-32},
          doi = {10.1051/aas:2000169},
archivePrefix = {arXiv},
       eprint = {astro-ph/0002122},
 primaryClass = {astro-ph},
       adsurl = {https://ui.adsabs.harvard.edu/abs/2000A&AS..143...23O}
}
\bibliographystyle{aasjournal}

\end{document}